\documentclass[
 preprint,
 amsmath,amssymb,
 aps,prf
]{revtex4-2}
\usepackage[table,xcdraw]{xcolor}
\usepackage[margin=1in]{geometry}
\usepackage{graphicx}
\usepackage{setspace}
\usepackage{xr}
\usepackage{hyperref}
\usepackage{todonotes}
\usepackage{subcaption}
\pdfoutput=1 
 \usepackage{booktabs}

\begin{document}
\title{Exploring regular and turbulent flow states in active nematic channel flow via Exact Coherent Structures and their invariant manifolds }\author{Caleb G. Wagner}
\affiliation{
Mechanical and Materials Engineering, University of Nebraska - Lincoln, Lincoln, NE 68588}
\author{Rumayel H. Pallock}
\affiliation{
Mechanical and Materials Engineering, University of Nebraska - Lincoln, Lincoln, NE 68588}
\author{Michael M. Norton}
\affiliation{
School of Physics and Astronomy, Rochester Institute of Technology, Rochester, NY 14623
}
\author{Jae Sung Park}
\affiliation{
Mechanical and Materials Engineering, University of Nebraska - Lincoln, Lincoln, NE 68588}

\author{Piyush Grover}
\email{piyush.grover@unl.edu}
\affiliation{
Mechanical and Materials Engineering, University of Nebraska - Lincoln, Lincoln, NE 68588}
\date{\today}
\begin{abstract}
This work is a unified study of stable and unstable steady states of 2D active nematic channel flow using the framework of Exact Coherent Structures (ECS). ECS are stationary, periodic, quasiperiodic, or traveling wave solutions of the governing equations that, together with their invariant manifolds, organize the dynamics of nonlinear continuum systems. We extend our earlier work on ECS in the preturbulent regime by performing a comprehensive study of stable and unstable ECS for a wide range of activity values spanning the preturbulent and turbulent regimes. In the weakly turbulent regime, we compute more than 200 unstable ECS that co-exist at a single set of parameters, and uncover the role of symmetries in organizing the phase space geometry. We provide conclusive numerical evidence that in the preturbulent regime, generic trajectories shadow a series of unstable ECS before settling onto an attractor. Finally, our studies hint at shadowing of quasiperiodic type ECS in the turbulent regime.

\end{abstract}
\maketitle

\section{Introduction}
Active fluids are viscous suspensions of constituents that consume chemical energy and convert it into mechanical work by generating stresses on the microscale, resulting in induced flows \cite{ramaswamy2019active}. They are governed by fully nonequilibrium dynamics \cite{kurzthaler2023out}, and their emergent spatiotemporal structures span multiple scales. In recent years, model active fluids have forged an unprecedented link between physics and biology \cite{balasubramaniam2022active}. Given their unique nature, there are several outstanding challenges in characterizing and controlling active fluids. Among various types of active fluids, \emph{active nematics} \cite{doostmohammadi2018active,mestre2022colloidal} --  suspensions of active, rod-like, and apolar components -- are of particular contemporary interest. This interest is in part motivated by the potential of exploiting the rich phenomenology of active nematics, including spontaneous coherent flows, dynamical vortex patterns, and chaotic hydrodynamics (i.e., low Reynolds number `active/mesoscale turbulence') \cite{alert2022active,martinez2019selection}, for design of smart materials. 

Extensile active nematics are inherently unstable to bend fluctuations \cite{Simha2002}. In the absence of boundaries or other influences, active nematic systems exhibit chaotic flows driven by topological defects \cite{Sanchez2012,thampi2014instabilities,decamp2015orientational,shankar2018defect}.  While the details vary from model to model, the general picture is that small distortions to a uniformly aligned active nematic system grow through hydrodynamic feedback, fueled by active stresses \cite{baskaran2009statistical}. When these distortions saturate, they create pairs of $\pm\frac{1}{2}$ defects, named according to their winding number. At steady state, these defects nucleate, self-propel, and annihilate, all while creating vortices with a correlation length $L_{\alpha}$. As the activity is increased, defect density increases and the size of vortices concomitantly decreases as $L_{\alpha}\sim\sqrt{K/{\alpha}}$, where $K$ is the elastic constant and $\alpha$ is the strength of the active stress such that $\sigma_{\alpha}\propto\alpha\mathbf{Q}$ \cite{doostmohammadi2018active}, with $\mathbf{Q}$ being the nematic alignment tensor.

Experimental efforts to modify the chaotic flows of active nematic system have utilized `soft' anisotropic substrate friction \citep{guillamat2016control,guillamat2017taming}, light-activated activity patterning \cite{ross2019controlling, zhang2021spatiotemporal, zarei2023light}  or confinement \citep{hardouin2019reconfigurable,opathalage2019self}. Given the importance of confined geometries in living active systems, the role of geometric confinement on dynamics has been explored computationally under a variety of geometries. A key control parameter in moderating the transition from laminar to chaotic flows is the domain size relative to $L_{\alpha}$. The ratio of this intrinsic length scale to imposed geometric length scales, such as channel width or domain radius, largely determines the spatiotemporal dynamics of the director and flow fields, and has been extensively explored experimentally and numerically \citep{giomi2012banding,shendruk2017dancing, norton2018insensitivity, chen2018dynamics, doostmohammadi2018active, opathalage2019self}.

In two-dimensions, the general trend \cite{samui2021flow,thampi2022channel} is that confining geometries much smaller than $L_{\alpha}$ result in stationary systems while systems much larger than $L_{\alpha}$ result in chaotic flows or mesoscale turbulence. Between these extremes, systems exhibit a variety of geometry-dependent dynamical states, such as coherent flows accompanied by either stationary or nontrivial periodic dynamics of the director field. For example, in the channel geometry, the so-called `dancing' state exists in this regime and is characterized by braiding orbits of  $+\frac{1}{2}$ defects and stationary $-\frac{1}{2}$ defects that decorate the walls of the channel \citep{shendruk2017dancing,chen2018dynamics,joshi2023dynamics}.

From an engineering perspective, there is substantial interest in understanding how to navigate the large space of spatiotemporal structures in confined active nematics \cite{norton2020optimal,shankar2022optimal,shankar2022spatiotemporal,mozaffari2021defect,falk2021learning}, by spatial, or spatiotemporal modulation of activity. There are also fundamental unanswered questions related to active turbulence in confinement \cite{thampi2022channel,henshaw2023dynamic}: how confined active fluids become turbulent, how to characterize them, and how to promote or inhibit turbulence. Some recent theoretical and computational studies have derived coarse-grained statistical descriptions \citep{alert2022active,alert2020universal,Giomi2015,linkmann2019phase,linkmann2020condensate} of active turbulence, focusing on energy spectra and spatial organization of statistically steady states, rather than deterministic dynamics. These studies have revealed that in contrast with inertial turbulence, energy transfer across scales is absent in active nematic turbulence. Furthermore, there exist universal scaling regimes in active nematic turbulence. While providing valuable information about the energetics and scaling properties of fully developed active turbulence, such studies do not provide much insight into developing capabilities for prediction or control of active turbulence, especially in confinement.

An alternative approach to the problem of turbulence is based on dynamical systems theory. The core premise, as originally applied to inertial turbulence, goes back to \citep{hopf1948mathematical,ruelle1971nature}. In this approach, one considers the fluid to be a deterministic dynamical system evolving in an infinite dimensional phase space \citep{cvitanovic2013recurrent}. The governing equations of a given hydrodynamic system can be written as an abstract first-order ODE \begin{equation}\dot{X}\mathop{=}F(X),\label{eq:abs}\end{equation} where $X$ denotes the $N$-dimensional state of the system. In computations, $N$ is a large but finite number corresponding to the degrees of freedom of the discretized system. The associated flow map is $f^t(X_0)\mathop{=}X_0+\int_{0}^tF(X(\tau))d\tau$, where $X_0$ is the initial condition. 

The dominant flow patterns of this system are understood in terms of \emph{Exact Coherent Structures} (ECS) \cite{graham2020exact,cvitanovic2005chaos} and the dynamical pathways connecting them. An ECS is a (stable or unstable) equilibrium, (time-) periodic, relative (time-) periodic, quasi-periodic or traveling wave solution of Eq. \ref{eq:abs}. Each type of ECS can be defined as a solution to a corresponding fixed point equation (FPE). In phase space, equilbria are 0D manifolds (points), periodic orbits and travelling waves are 1D manifolds (curves), and relative periodic orbits are 2D manifolds (surfaces). The FPE for an equilibrium solution $X_{\text{eq}}$ is $F(X_{\text{eq}})\mathop{=}0$, while any point $X_{\text{P}}$ on a periodic orbit (PO) satisfies $f^{T}(X_{\text{P}})\mathop{=}X_{\text{P}}$, where $T$ is the time period. Similarly, a point $X_{\text{RP}}$ on a relative periodic orbit (RPO) satisfies $f^{T}(X_{\text{RP}})\mathop{=}\tau_{x}(\ell) X_{\text{RP}}$, where $\tau_x(\ell)$ is an operator that shifts the solution in the streamwise direction by a distance $\ell$. While stable ECS can be computed by time-dependent simulations of Eq. \ref{eq:abs}, the computation of unstable ECS requires the solution of the above mentioned FPEs.

The stability of each ECS can be determined by linearization about its trajectory in the phase space. In general, if a $d$-dimensional ECS in an $N-$dimensional phase space has $n_u$ linearly unstable directions forming the local unstable subspace, and $n_s$ linearly stable directions forming the local stable subspace, then $n_u+n_s+d=N$. Hence, away from bifurcation points, each unstable ECS behaves like a (high-dimensional) saddle. Associated to such an ECS are $(n_u+d)-$dimensional unstable manifolds and $(n_s+d)-$dimensional stable manifolds. Mathematically, the stable and unstable manifolds are sets of initial conditions that converge to the ECS in forward and backward time, respectively. These invariant manifolds are dynamical pathways connecting distant regions of phase space. 

The steady states that are directly observed in experiments or time-dependent numerical simulations correspond to various types of \emph{stable} ECSs. It has been conjectured that turbulent flows correspond to chaotic trajectories meandering through the phase space and visiting the neighborhoods of different ECS (all of which are \emph{unstable}) in a recurring fashion \citep{park2015exact,budanur2019geometry,suri2020capturing}. Therefore, the ECS and their invariant manifolds act as an organizing template for the complicated spatiotemporal motion of the fluid. This minimal description of complicated flows is well-suited for devising near-optimal control inputs using activity patterning. Such an approach has been successfully applied in several low-dimensional nonlinear dynamical systems \cite{koon2000heteroclinic,sakamoto2013case,flasskamp2012solving}, where (near-) optimal control trajectories have been computed solely based on the knowledge of various equilibria or periodic orbits, and their stable and unstable manifolds, without solving any formal optimization problem.

In our earlier work \cite{wagner2022exact}, we reported the co-existence of a chaotic attractor, multiple regular attractors (i.e., stable ECS), and a large number of unstable ECS in the phase space, at a single value of activity below the turbulent transition. This paper extends this previous analysis in several directions. In section \ref{sec:attractors}, we present a phase diagram constructed via a comprehensive study of attractors for a wide range of activity values. We find that above a critical activity value, there are no regular (i.e., non-chaotic) attractors in the system, and all trajectories are chaotic, signifying the transition to turbulence. Moving beyond attactors, we discuss the computation and classification of various stable and unstable ECSs in preturbulent and turbulent activity regimes in section \ref{sec:ECS}. We show that the explicit use of system symmetries enables the computation and classification of system states. The symmetry viewpoint \cite{golubitsky2003symmetry,hoyle2006pattern} also aids in visualization and understanding of the phase space geometry.

The conjecture \cite{cvitanovic2013recurrent} that turbulent trajectories shadow various ECSs, guided by the associated invariant manifolds and heteroclinic connections, has been confirmed in recent numerical \cite{krygier2021exact} and experimental \cite{crowley2022turbulence,suri2020capturing} studies in certain inertial turbulent systems. In section \ref{sec:transient-flows}, we review the tools introduced in these prior studies \cite{krygier2021exact, crowley2022turbulence}, and employ them to provide numerical evidence of shadowing of ECSs and their invariant manifolds in the active nematic system in the preturbulent regime. Finally, we also discuss our efforts to confirm shadowing of ECS in the turbulent regime in the same section.

\section{Model, Symmetries and Computational Framework} \label{sec:model}
Following earlier work \cite{Marenduzzo2007,Thampi2014,giomi2014defect,shendruk2017dancing}, we model the active nematic system in terms of the velocity $\mathbf{u}(\mathbf{r},t)\mathop{=}(u,v)$, and nematic alignment tensor $\mathbf{Q}(\mathbf{r},t)$. The symmetric and traceless $\mathbf{Q}$ tensor can be parameterized as $Q_{ij} = q (n_in_j-0.5\delta_{ij})$, where the scalar $q$ and unit vector $\hat{\mathbf{n}}$ describe the degree and direction of nematic ordering.
The governing equations are:\begin{align}\begin{split}
&\rho \left(\partial_t + \mathbf{u} \cdot \boldsymbol{\nabla}\right)\mathbf{u} = -\boldsymbol{\nabla} p + 2 \eta (\boldsymbol{\nabla} \cdot \mathbf{E}) - \alpha (\boldsymbol{\nabla} \cdot \mathbf{Q}), \\
&\left(\partial_t + \mathbf{u} \cdot \boldsymbol{\nabla}\right)\mathbf{Q} + \mathbf{W} \cdot \mathbf{Q} - \mathbf{Q} \cdot \mathbf{W} = \lambda \mathbf{E} +  \gamma^{-1} \, \mathbf{H},\\
 &\qquad \boldsymbol{\nabla} \cdot \mathbf{u} = 0,\end{split}\label{eq:main}
\end{align}
where
\begin{align}
E_{ij} &= \frac{1}{2} \left(\partial_i u_j + \partial_j u_i \right); \qquad W_{ij} = \frac{1}{2} \left(\partial_i u_j - \partial_j u_i \right)  \\
\mathbf{H} &= A \left[\mathbf{Q} \mathbin{-} b \mathrm{Tr}\left(\mathbf{Q}^2\right) \mathbf{Q}\right] \mathbin{+} K \nabla^2 \mathbf{Q}. \label{eq:H-intro}
\end{align}
 The response of the material to activity is captured in the Navier-Stokes equation as the active stress $-\alpha \mathbf{Q}$, which models an active particle as a force dipole \cite{Shaebani2020}. We do not include passive elastic stresses in this equation, since they are subdominant \cite{wagner2022exact,norton2018insensitivity,Koch2021}. In the $\mathbf{Q}$-equation, the l.h.s.  is the convective derivative of $\mathbf{Q}$ with respect to $\mathbf{u}$. In addition to the usual $(\mathbf{u} \cdot \boldsymbol{\nabla})$ contribution, there is also the commutator product $\mathbf{W} \cdot \mathbf{Q} - \mathbf{Q} \cdot \mathbf{W}$, which accounts for rotations of $\mathbf{Q}$ in the convected coordinates. On the r.h.s, $\lambda\mathbf{E}$ is the flow-alignment term, and $\mathbf{H}$ derives from a free energy functional that penalizes distortions and drives the creation of nematic order. In the present study, we neglect flow-alignment to focus on essential aspects of the system \cite{shankar2018defect,blow2017motility}. Other aspects of the nematohydrodynamic model, including generalizations and detailed physical interpretation, can be found in \cite{Shaebani2020}.
The domain is a periodic 2D channel, parameterized as $(x,y) \in \left[0, L \right] \times \left[0, h \right]$, with $x$ the periodic coordinate. The channel walls impose a no-slip boundary condition on $\mathbf{u}$ and strong perpendicular anchoring on $\mathbf{Q}$.

For our simulations, we adopt the non-dimensionalization used by Ref. \cite{Koch2021}, where the fundamental length and time scales $\ell, \tau$ are defined as
\begin{equation}
\ell = \frac{K^{1/2}}{A^{1/2}}, \qquad \tau = \frac{\gamma}{A} \label{eq:ellandtau-definition}
\end{equation}
The corresponding velocity scale $v_0$ is
\begin{equation}
v_0 = \frac{\ell}{\tau} = \frac{K^{1/2}}{A^{1/2}} \cdot \frac{A}{\gamma} = \frac{K^{1/2} A^{1/2}}{\gamma} 
\end{equation}
In these units, equations \eqref{eq:main} become
\begin{align}
&\mathrm{Re}_{\mathrm{n}}\left(\partial_t + \mathbf{u} \cdot \boldsymbol{\nabla}\right)\mathbf{u} = -\boldsymbol{\nabla} p + 2 (\boldsymbol{\nabla} \cdot \mathbf{E})  - \frac{\mathrm{R}_{\mathrm{a}}}{\mathrm{Er}} \left( \boldsymbol{\nabla} \cdot \mathbf{Q} \right), \nonumber \\
&\left(\partial_t + \mathbf{u} \cdot \boldsymbol{\nabla}\right)\mathbf{Q} + \mathbf{W} \cdot \mathbf{Q} - \mathbf{Q} \cdot \mathbf{W} = \lambda \mathbf{E} +  \mathbf{H}, \label{eqs:main}\\
 &\qquad \boldsymbol{\nabla} \cdot \mathbf{u} = 0. \nonumber \\
 &\mathbf{H} = \mathbf{Q} \mathbin{-} b \mathrm{Tr}\left(\mathbf{Q}^2\right) \mathbf{Q} \mathbin{+} \nabla^2 \mathbf{Q},
\end{align}
where
\begin{align}
\mathrm{Re}_{\mathrm{n}} &= \frac{\rho v_0 \ell}{\eta}, \\
\mathrm{Er} &= \frac{\eta v_0 \ell}{K} = \frac{\eta}{K} \cdot \frac{K}{\gamma} = \frac{\eta}{\gamma}, \\
\mathrm{R}_{\mathrm{a}} &= \frac{K}{A} \cdot \frac{\alpha}{K} = \frac{\alpha}{A}. \label{eq:Ra-definition}
\end{align}
$\mathrm{R}_{\mathrm{a}}$ is the dimensionless activity, $\mathrm{Er}$ is the Ericksen number, and $\mathrm{Re}_{\mathrm{n}}$ was identified in  Ref. \cite{Koch2021} as the microscopic Reynolds number. Consistent with prior work \cite{shendruk2017dancing,doostmohammadi2017onset}, we fix the parameters $\mathrm{Er=1} \text { and } \mathrm{Re_n}=0.0136$. The channel dimensions are $L \mathop{=} 80$ and $h \mathop{=} 20$ in the dimensionless units.
\subsection{Symmetries}
   A dynamical system of the form given by Eq.~\eqref{eq:abs} is said be equivariant under a group action $\sigma$ if the time evolution and the group operation commute, i.e. for every state $X$ and time $T$, the relationship $\sigma(f^T(X))=f^T(\sigma(X))$ is satisfied. The system of Eqs.~\eqref{eq:main} in a periodic channel is equivariant under the continuous one-parameter group of $x$ translations, $\tau_x(\ell)$, as well as the $x$ and $y$ reflections, denoted $\sigma_x$ and $\sigma_y$, where
\begin{align} \sigma_x[u,v,Q_{11},Q_{12}](x,y)=[u,-v,Q_{11},-Q_{12}](x,h-y),\nonumber \\
\sigma_y[u,v,Q_{11},Q_{12}](x,y)=[-u,v,Q_{11},-Q_{12}](L-x,y),\nonumber\\
\tau_{x}(\ell)[u,v,Q_{11},Q_{12}](x,y)=[u,v,Q_{11},Q_{12}](x+\ell,y),
\end{align}
and $\ell$ is any real number. We will use the notation $T_k$ to denote the discrete $k$-fold translation group with operation $\tau_x(\frac{L}{k})$. We use the following convention while naming various ECS: `T$k/m$' refers to an ECS with $T_k$ symmetry, and the integer $m$ is an identifier used to differentiate between different ECS with the same symmetry.
Suppose an initial condition $X_0$ is symmetric with respect to any of the above group operations or their compositions, e.g., let $X_0=\sigma_x(X_0)$. Then, the future states $f^T(X_0)$ of the system will retain the same symmetries \cite{cvitanovic2005chaos}, and hence, $f^T(X_0)=\sigma_x(f^T(X_0))$ for all $T>0$. In that case, the state $X_0$ and all its future iterates are said to belong to the $\tau_x$ symmetry subspace $Fix(\sigma_x) \subset \mathbb{R}^N$, where $Fix(\sigma_x)=\{X\in\mathbb{R}^n|\sigma_x(X)=X\}$. Our discussions in the following sections will reveal that these symmetries are a powerful tool for finding and categorizing various ECSs, and analyzing the phase space geometry.

\subsection{Computational Framework}
We have developed an open-source computational toolbox titled `Exact Coherent Strucures in Active Matter' (ECSAct)\cite{ecsact} using the open-source pseudospectral code Dedalus \citep{burns2020dedalus}. 
Dedalus is a Python-based, MPI-parallelized general purpose solver for initial-value, boundary-value, and eigenvalue problems on spectrally representable domains. Nearly arbitrary equations can be efficiently solved via state-of-the-art algorithms that exploit sparsity, arithmetic trees, and transforms. Pseudospectral methods are well-suited for solving nonlinear PDEs in simple geometries \citep{Boyd}. For channel geometries, Dedalus implements a Fourier basis for the periodic directions and Chebyshev polynomials for the wall-normal direction. 

While Dedalus can natively solve time-dependent problems, it currently lacks sophisticated solvers that can handle highly nonlinear multi-dimensional FPEs such as the ones that need to be solved to compute unstable ECSs. To solve such FPEs, ECSAct uses modified Newton-Raphson algorithms \citep{viswanath2007recurrent}. Two key ingredients are adaptive `hookstep' step-size selection to improve global convergence \citep{Dennis1996}, and a matrix-free GMRES \citep{saad1986gmres,chandler2013invariant} algorithm for solving the linear BVP at each iteration. The matrix-free methods are essential because they scale efficiently to the large problem dimensions encountered in hydrodynamic simulations. ECSAct also contains routines for linear stability analysis of the various types of ECSs, as well as for performing symmetry operations such as projections into a symmetric subspace, etc. More details on these algorithms and implementation of ECSAct can be found in the ECSAct documentation available on the Github repository \cite{ecsact}.

All ECS and heteroclinic connections reported here were computed using $128$ Fourier modes and $32$ Chebyshev modes, corresponding to a phase space dimension of $\sim 4 \mathbin{\times} 128 \mathbin{\times} 32 \mathop{=} 16384$.

\section{Results}
 
\subsection{Attractors} \label{sec:attractors}
\subsubsection{Search strategy}
Nonlinear dynamical systems can have multiple attracting states. To discover an attractor by forward time integration, the initial conditions must be within the corresponding basin of attraction. However, for multiple attractors, there is no general method for ensuring that all basins of attraction have been sampled. Thus, heuristic methods are usually required. In this section, we explain the heuristics we have used for mapping out attractors in AN channel flow. 

As a starting point, we refer to our previous work \cite{wagner2022exact} on AN channel flow at a single value of activity. There, we found three coexisting attractors: three- and four-fold vortex lattice (`dancing disclinations') PO states, and a localized chaotic attractor. Because all three were well-separated from each other in phase space, a cursory search by forward time integration could easily have overlooked one by failing to provide initial conditions in each basin of attraction. Finding all three required a more principled approach, in which initial conditions were chosen in a set of discrete symmetry subspaces. The idea is that the dynamics tends to prefer configurations that match the system's length scales, corresponding to flow states with exact or approximate discrete translational symmetry. In a channel with aspect ratio width/height $\approx 4$, it is plausible that the non-chaotic attractors have 3, 4 or 5 fold translational symmetry: for example, a row of 3, 4 or 5 identical (or nearly so) vortices. Indeed, this configuration coincides with the stable vortex lattice PO consistently observed in previous studies. This physical reasoning concurs with the more general dynamical systems context, where the role of symmetry subspaces in organizing the phase space geometry is well-established. 
	\begin{figure}
    \includegraphics[width=0.99\textwidth]{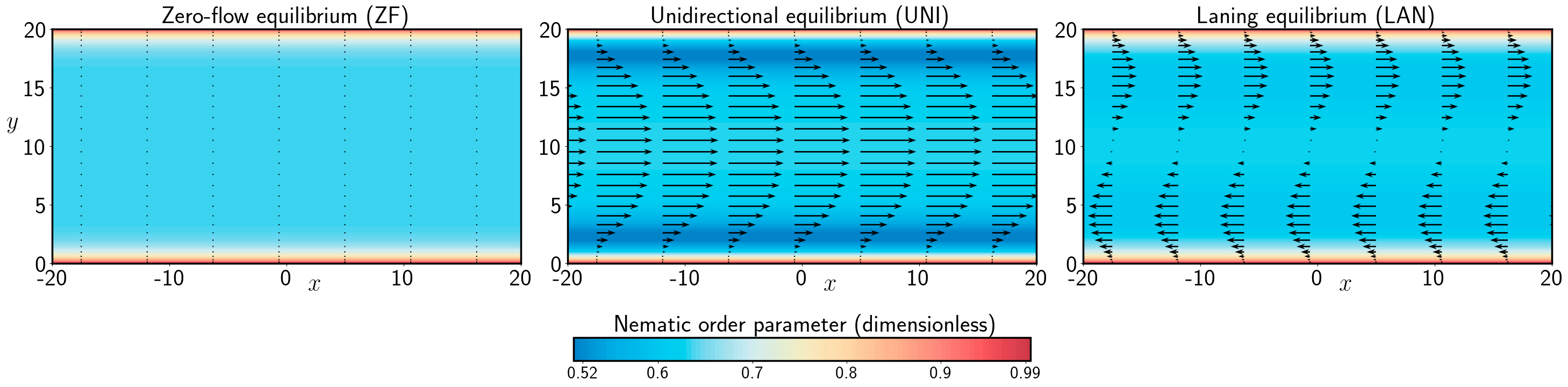}
    \caption{The three equilibria with zero or 1D flow for $\mathrm{R_a}= 3.4$.}
        \label{fig:1deq}
    \end{figure}
Our strategy, therefore, is to search for attractors on or near symmetry subspaces that match the system's preferred length scales. One method is to choose initial conditions in a symmetry subspace and explicitly enforce that symmetry under forward time integration. However, doing so would overlook key features that are only represented in the full phase space. These include dynamical connections between ECS in different symmetry subspaces, as well as any attractors that are small perturbations from a symmetric state, e.g., a vortex lattice with one of the vortices slightly distorted. Therefore, we perform our searches for attractors in the full phase space.
\begin{table}
\centering
\begin{tabular}{p{10mm}>{\centering\arraybackslash}p{15mm}>{\centering\arraybackslash}p{10mm}>{\centering\arraybackslash}p{10mm}>{\centering\arraybackslash}p{15mm}>{\centering\arraybackslash}p{15mm}>{\centering\arraybackslash}p{15mm}>{\centering\arraybackslash}p{15mm}}  \hline \hline \addlinespace[2mm]
$n_x$ & $\text{Re}(\lambda)$ & & $n_x$ & $\text{Re}(\lambda)$& & $n_x$ & $\text{Re}(\lambda)$ \\
\hline
\addlinespace[2mm]
6 & 0.623 &    & 10 & 0.372 &    & 9  & 0.207\\ 
7 & 0.603 &    & 3  & 0.296 &    & 10 & 0.137\\ 
5 & 0.595 &    & 11 & 0.254 &    & 12 & 0.12\\ 
8 & 0.55  &    & 7  & 0.249 &    & 5  & 0.115\\ 
4 & 0.497 &    & 8  & 0.246 &    & 11 & 0.042\\ 
9 & 0.471 &    & 6  & 0.208 &    & 2  & -0.021\\ 
\addlinespace[1mm]
\hline \hline            
\end{tabular}
\caption{\footnotesize Features of the leading eigenvectors of the dynamics linearized about UNI at $\mathrm{R_a}=3.4$. The eigenvectors are pure Fourier modes in $x$, of the form $e^{\pm 2\pi n_x/L}g(y)$, where $L=80$ is the channel width. $\lambda$ is the eigenvalue, so positive (negative) $\text{Re}(\lambda)$ means the eigenvector represents an unstable (stable) direction.}
\label{table:UNI_eigenvectors}
\end{table}
In practice, the set of equilibria with 1D flow, henceforth referred to as `1D equilibria', shown in Fig. \ref{fig:1deq} serve as convenient `entry points' to the symmetry subspaces. These three equilibria \cite{duclos2018spontaneous,Walton2020,wagner2022exact} have continuous translational symmetry, so they belong to every discrete translational symmetry subspace. For the same reason, the unstable eigenvectors of the system linearized about these equilibria are of the form $e^{\pm 2\pi n_x/L}g(y)$ and therefore possess discrete translational symmetry. These symmetries also tend to be the most dynamically relevant ones, as they are precisely those that grow and persist as a trajectory leaves the equilibrium. By contrast, the dynamically unfavorable symmetries tend to be represented by the stable eigenvectors, which are damped out in the neighborhood of the equilibrium. Table \ref{table:UNI_eigenvectors} illustrates this idea for UNI at $\mathrm{R_a} = 3.4$. In this case, the unstable eigenvectors have wavelengths ranging from one-third to one-twelfth of the channel width $L$, corresponding to discrete symmetries $T_3, T_4, \ldots , T_{12}$. As demonstrated in the following sections, these symmetries are indeed important for characterizing typical time-dependent trajectories, while higher order symmetries, $T_{n>12}$, are inconsequential.

In the end, our search consisted of the following steps:
\begin{enumerate}
\item For each $\mathrm{R_a} \in \{0.1 \, n, n = 1 \ldots 50\}$, compute the unstable eigenvectors of UNI, LAN, and ZF.
\item For each unstable eigenvector $\mathbf{v}$ construct the initial condition $\mathbf{X}_0 + \epsilon (\mathbf{v}/||\mathbf{v}||)$, where $\mathbf{X}_0$ is one of UNI, LAN, or ZF; and $\epsilon = 10^{-5}$. From this initial condition, run a time-dependent simulation for $10^4 \tau$ time units. Denote the final state $\mathbf{X}_{\infty}$.
\item Run the Newton--Raphson solver with $\mathbf{X}_{\infty}$ as the initial guess. If the solver converges, compute the stability and see if it is an attractor.
\end{enumerate}
We also performed branch continuation for several ECS, checking whether an unstable solution branch became stable at higher or lower activity. However, most attractors were found using the time-dependent method described above.

\subsubsection{Description and Classification}
\begin{figure}
      \includegraphics[width=0.99\textwidth]{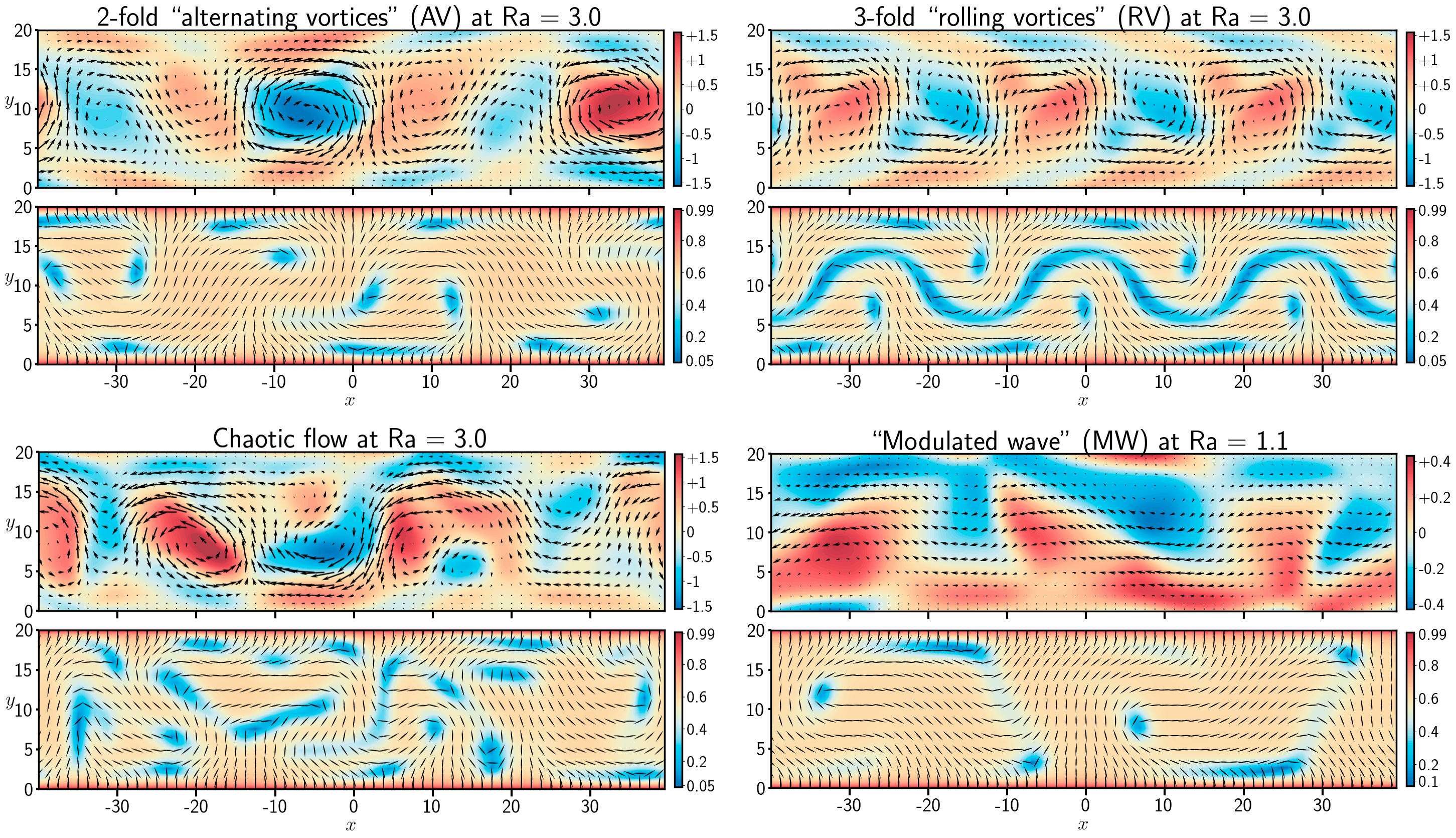}
    \caption{\footnotesize Snapshots of some of the attractors: The vortical flows (top row) and chaotic flow (bottom left) are at $\mathrm{R_a}=3.0$ and are shown with the same colorbar scales. The modulated wave (bottom right) is at $\mathrm{R_a}=1.1$; due to the smaller activity, it is shown with different colorbar scales. The alternating vortex lattice RPO and the modulated wave RPO both have shift-reflect symmetry $\sigma_xT_2$, the rolling vortex lattice RPO has three-fold translation symmetry $T_3$, while the chaotic flow does not have any symmetry.}
        \label{fig:att_flow}
    \end{figure}
\begin{figure}
    \includegraphics[width=0.95\textwidth]{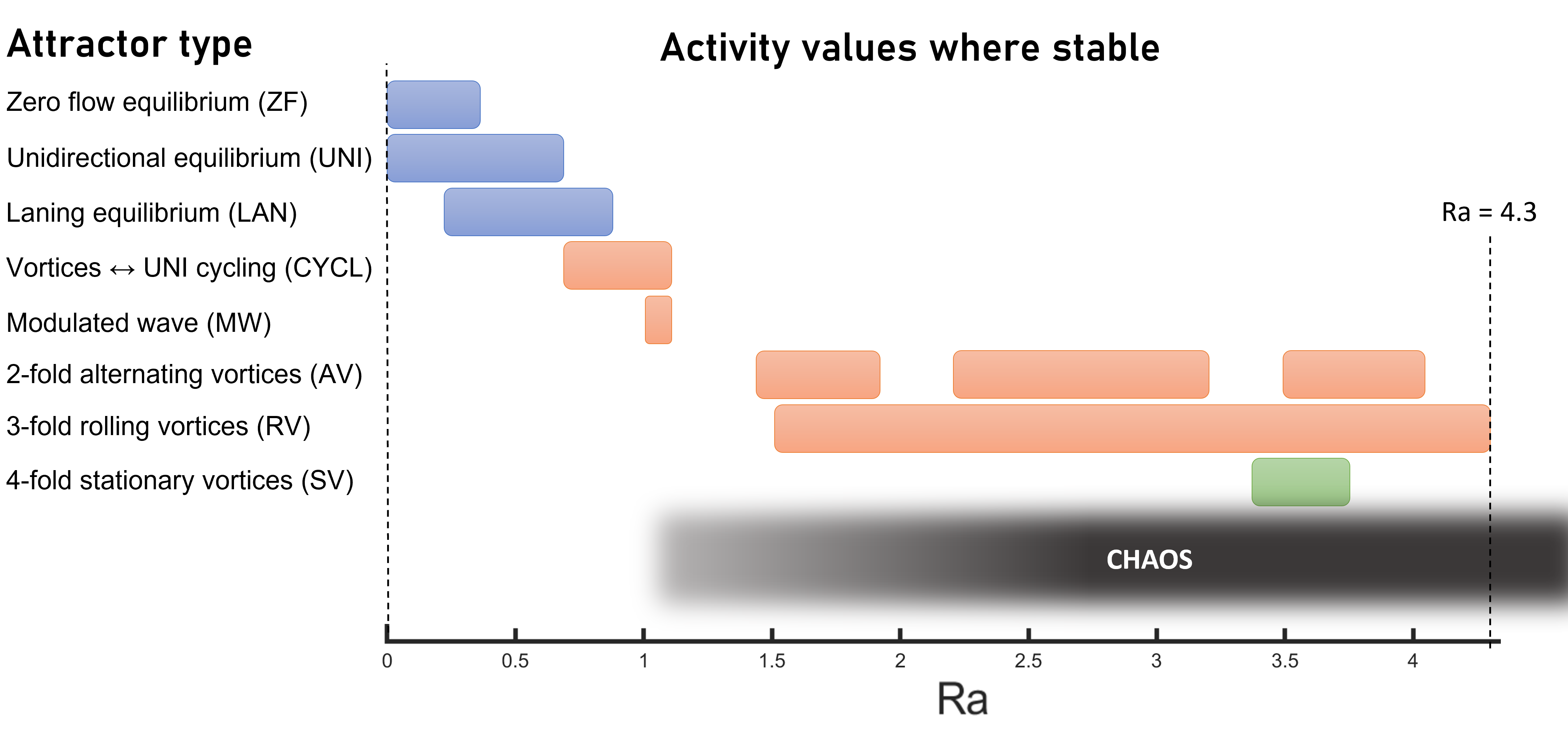}
    \caption{\footnotesize Attractors as a function of activity $\mathrm{R_a} \equiv \alpha/A$ (Eq.~\ref{eq:Ra-definition}). The channel dimensions are $80 \ell \times 20 \ell$, where $\ell$ is defined from the material constants of the nematic (Eq.~\ref{eq:ellandtau-definition}). The flow alignment $\lambda$ is $0$. The horizontal bars group the attractors by symmetries and other qualitative flow characteristics. A blue color indicates an equilibrium, orange an RPO, and green a PO. Some of the bars are composed of more than one ECS; for example, there are two qualitatively similar 3-fold rolling vortices between $\mathrm{R_a} \!=\! 3.34$ and $\mathrm{R_a} \!=\! 3.86$. Real-space flow snapshots are shown in Fig.~\ref{fig:att_flow}, and the full dataset of ECS is given in the Supplementary material \cite{suppdata2023}.}
        \label{fig:att}
    \end{figure}

The non-chaotic attractors can be categorized by the symmetries of Eq. \ref{eq:main}. For $0 < \mathrm{R_a} < 0.7$, one or more of the 1D equilibria (ZF, UNI, LAN) is stable. The equilibria or POs borne as a result of bifurcations from these attractors break the continuous translational symmetry, while picking a discrete translational symmetry. The \emph{stable} ECS (i.e., the attractors) will be the ones that pick up symmetries with dynamically favorable length scales, which in turn depend on the system's intrinsic length scales. For $\mathrm{R_a}\approx 0.7$, where UNI becomes unstable, an RPO with $T_4$ symmetry is born through a homoclinic bifurcation. This RPO cycles between the neighborhoods of the UNI equilibrium and the PO corresponding to a 4-fold vortex lattice state. The next attractor to appear (near $\mathrm{R_a} \approx 1.1$) is an RPO invariant with respect to $\sigma_x T_2$, which describes a \emph{shift-reflect} symmetry \cite{gibson2008visualizing}. Though this particular attractor (labeled `modulated wave' (MW)) occupies a narrow range of activity (5th row in Fig. \ref{fig:att}), the $\sigma_x T_2$ symmetry appears repeatedly in other attractors all the way to $\mathrm{R_a} \approx 4.1$. A snapshot of a MW RPO is shown in Fig. \ref{fig:att_flow} (bottom right). 

For the channel dimensions and material constants considered here, the most robust attractors have a $T_3$ symmetry. Although there is a stable PO with $T_4$ symmetry (green bar in Fig. \ref{fig:att}), it is stable over a relatively small range of activity, and even then only weakly so.

The symmetries loosely correlate with additional flow characteristics. In particular, the $T_n$ symmetries are associated with one of the following: (a) cycling between a nearly unidirectional flow and a vortex lattice PO (4th row in Fig. \ref{fig:att}); (b) left- or right-drifting vortex lattice RPOs (labeled `rolling' vortex lattice, row 7); or (c)  vortex lattice PO (row 8). In cases (b) and (c), the vortices are co-rotating. A snapshot of a rolling vortex lattice RPO is shown in Fig. \ref{fig:att_flow} (top right). In contrast, because $\sigma_x$ switches the sign on the vorticity, the shift-reflect symmetry ($\sigma_x T_2$) tends to produce a pair of counter-rotating vortices (row 6). A snapshot of this RPO, labeled `alternating' vortex lattice, is shown in Fig. \ref{fig:att_flow} (top left). 

We did not find any non-chaotic attractors in the activity range $1.1\leq \mathrm{R_a} \leq 1.4$ for the zero flow alignment case considered in this work. By repeating similar calculations for several non-zero values of flow-alignment, we verified that generically, at least one non-chaotic attractor can be found for all values of activities below the turbulent transition.
\subsubsection{Chaotic attractors}
\begin{figure}
    \includegraphics[width=0.41\textwidth]{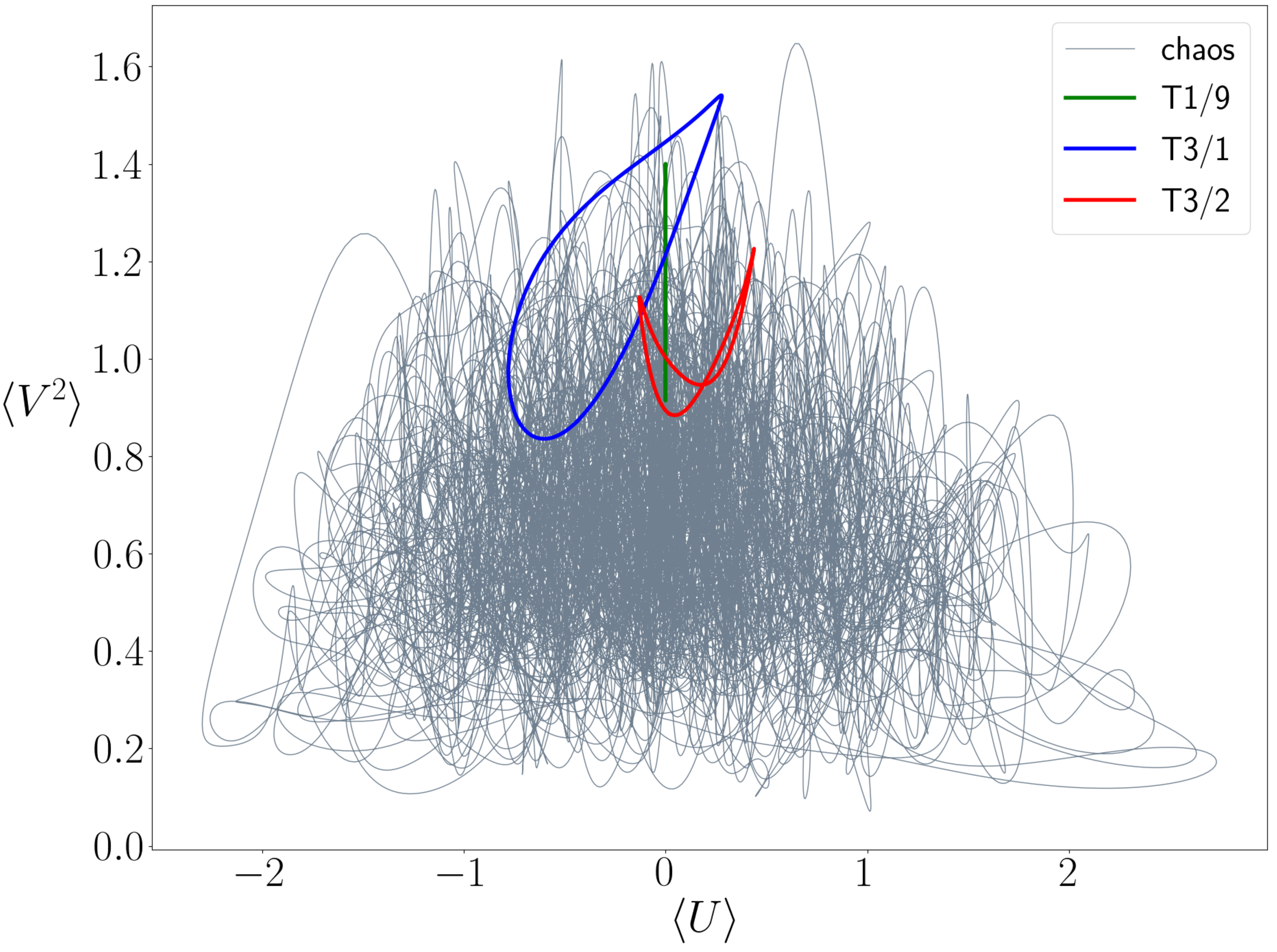}
    \hspace{.3in}
    \includegraphics[width=0.495\textwidth]{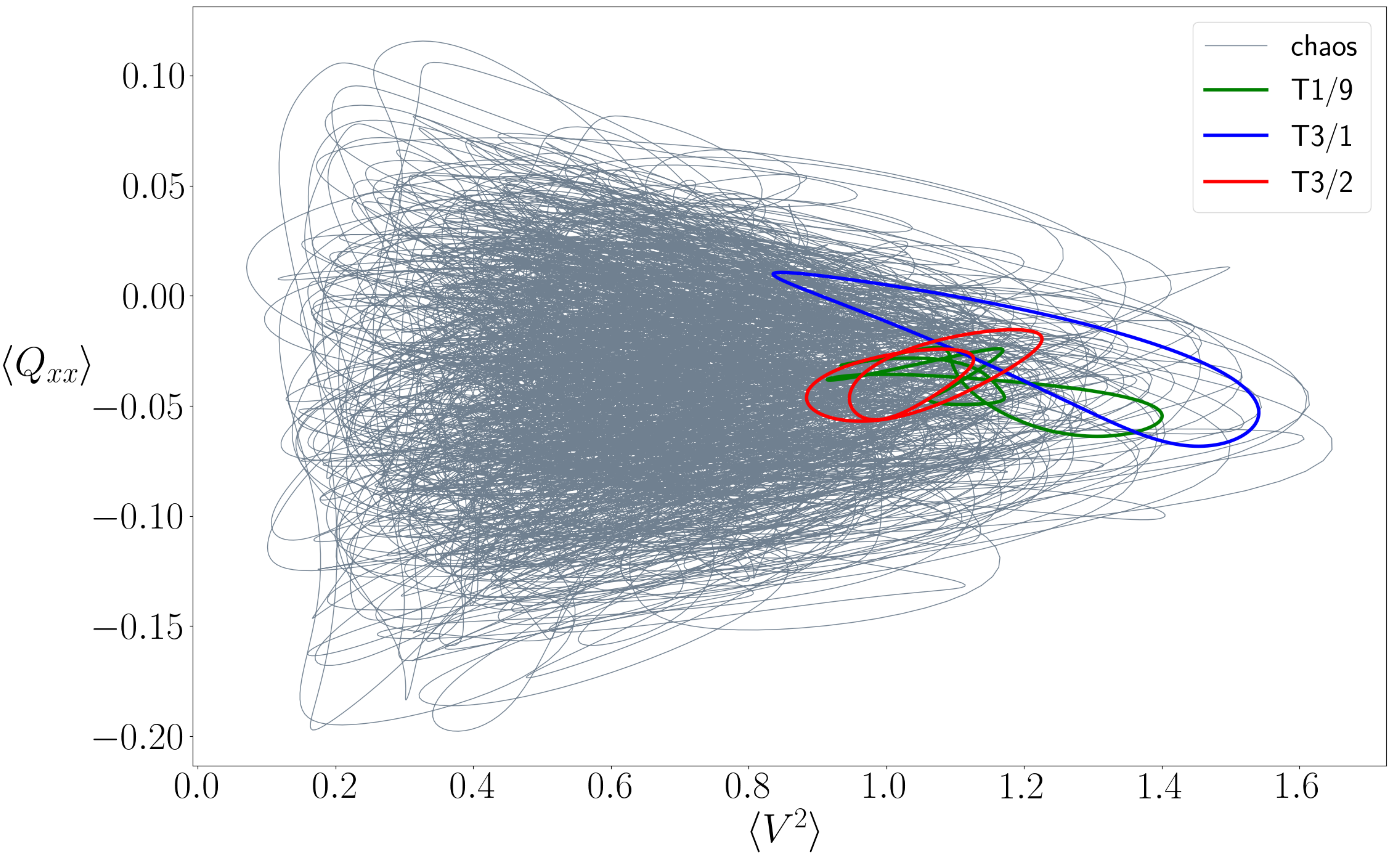}
    \caption{\footnotesize A chaotic trajectory and the three attractors projected on the $(\langle U\rangle,\langle V^2 \rangle)$ (left) and $(\langle V^2 \rangle,\langle Q_{xx}\rangle)$ (right) planes, for $\mathrm{R_a}=3.4$.}
        \label{fig:edge_R34}
\end{figure}

Surprisingly, there is evidence of chaotic attractors for activity as low as $\mathrm{R_a} \approx 1.1$. By contrast, stable ECS exist all the way up to $\mathrm{R_a} \approx 4.3$. In between, the phase space can be partitioned into basins of attraction for the stable ECS and any chaotic attractors. The former consists of all points in phase space that converge to a stable ECS as $t \rightarrow +\infty$, and the latter consists of all points that converge to a chaotic attractor as $t \rightarrow +\infty$. For visualization purposes, we project trajectories on to a reduced 3D phase space $(\langle u\rangle,\langle v^2\rangle,\langle Q_{11}\rangle)$ \cite{wagner2022exact}, where $\langle . \rangle$ denotes the instantaneous channel average. Visual inspection suggests that chaotic trajectories in this phase space projection are localized to a region separate from the stable ECS, see Fig. \ref{fig:edge_R34} for the case of $\mathrm{R_a}=3.4$. The symmetry-reduced distance (Eq. \ref{eq:symmetry-reduced-distance}) between a chaotic trajectory and the attractors at $\mathrm{R_a}=3.4$ was found to be $O(1)$, providing further proof of our assertion. 

Interestingly, the unstable manifold of the unidirectional equilibrium appears to play a role in dividing these two regions. In our previous work \cite{wagner2022exact}, we found robust heteroclinic connections between UNI and certain stable ECS, as well as trajectories connecting UNI to the chaotic attractor. Overall, this phase space partitioning is reminiscent of the subcritical transition to turbulence in pipe flow \cite{pringle2012minimal,duguet2013minimal,park2015exact}. These studies posit that a codimension-1 manifold separates the basins of attraction of the laminar and the chaotic states. Such a manifold corresponds to the stable manifold of an unstable equilibria (or travelling wave) with one unstable direction. The question of existence and characterization of a similar edge manifold in the phase space of active nematic channel flow is left for future work. 

\subsection{Unstable ECS}\label{sec:ECS}
In addition to the regular (stable ECS) and chaotic attractors, the phase space is populated by several unstable ECSs. The unstable ECSs are coherent steady states that are not \emph{directly} observed in experiments or time-dependent numerical simulations. Our interest in computing and characterizing these states is motivated by their crucial role organizing the global dynamics in both the non-turbulent and turbulent regime.

\begin{figure}
        \includegraphics[width=0.99\textwidth]{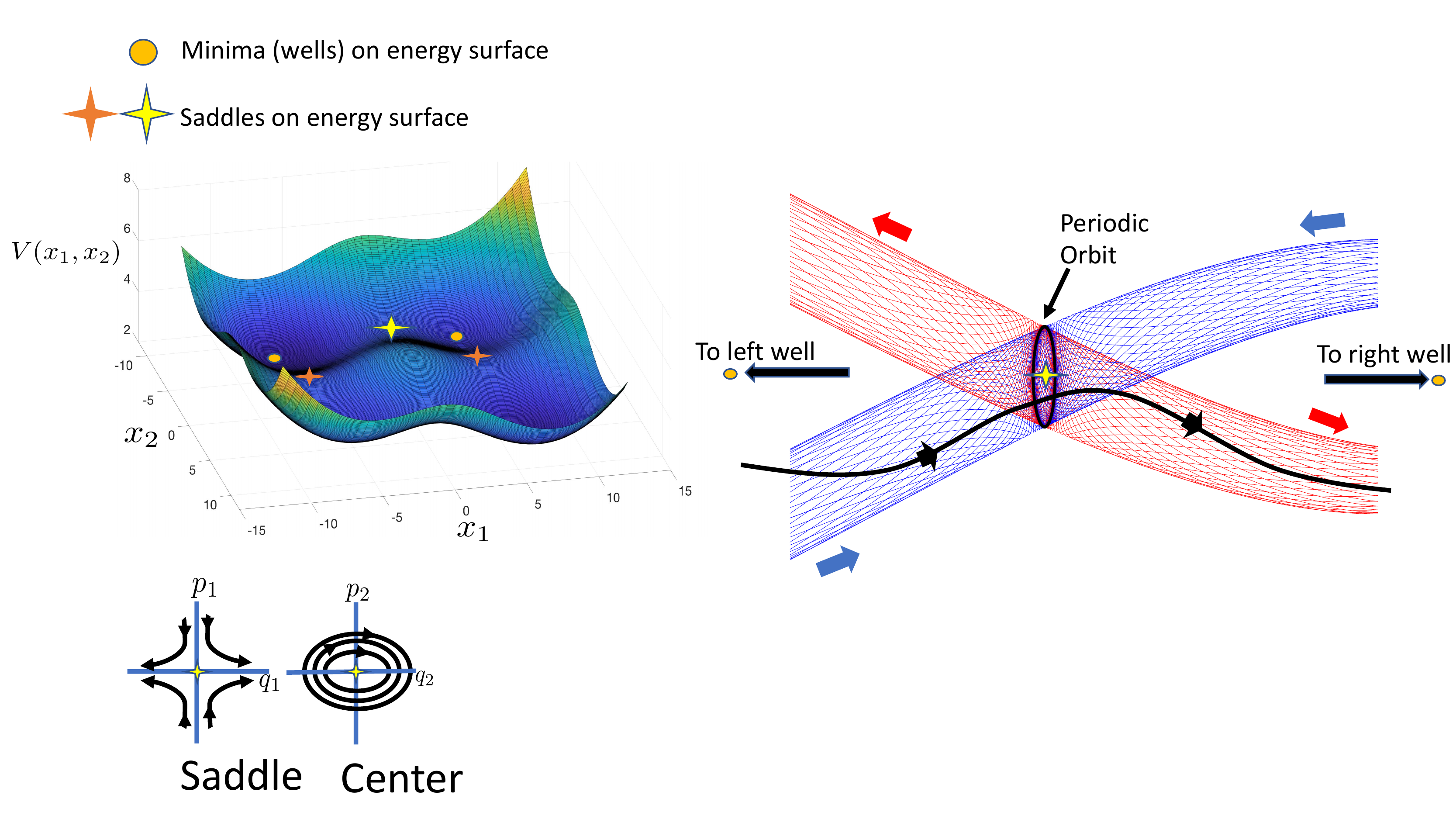}
    \caption{\footnotesize Left: (Top) Potential energy surface of a 2 DOF conservative dynamical system with two minima and three saddles highlighted. (Bottom) The phase space near the energy surface saddle can be locally described as a direct product of a saddle fixed point and a center fixed point on orthogonal planes. Right: Periodic orbit around the saddle in yellow has 2D stable and unstable manifolds. A trajectory transiting from the left potential well to the right potential well travels inside the stable manifold towards the PO, and inside the unstable manifold while moving away from the PO. Hence, the PO here acts like an extended saddle point.}
        \label{fig:2denergy}
\end{figure}

In several classes of dynamical systems, the natural dynamical pathways between a pair of stable steady states are often found to be mediated by \emph{unstable} steady states. Consider the recent experimental study of the dynamics of a ball rolling on a 2D multi-well surface \cite{ross2018experimental}. This conservative two degree-of-freedom system has a 3D phase space for each fixed value of total energy. A trajectory can transition between a pair of two wells (potential energy minima) via a bottleneck that opens up at an energy level above that of the intermediate saddle point of the energy surface, see Fig. \ref{fig:2denergy}. In addition to this energetic requirement however, the desired transition can happen only if the trajectory lies inside the 2D cylindrical stable manifold of the \emph{unstable} periodic orbit that exists around the saddle point of the energy surface. If this latter condition is met, the trajectory approaches the saddle point region while travelling inside the stable manifold, and is eventually transferred to the second well via the unstable manifold of the same periodic orbit. If the condition is not met, the trajectory is reflected back to first the well, even if it satisfies the energetic requirement. 

The above mentioned example highlights the role of invariant manifolds of saddle type (unstable) steady states in determining global phase space transport. Similar results have been obtained in orbital mechanics \cite{koon2000heteroclinic,grover2009designing}, ship dynamics \cite{naik2017geometry}, beam buckling \cite{zhong2018tube}, and solitary wave propagation \cite{mohammed2022phase}, etc. This body of work suggests that the unstable ECSs are expected to play a similarly important role in energy efficient dynamical pathways that transfer the system between various attractors in the preturbulent regime. The geometry of phase space can also assist in the discovery of new flow states that may become stable at different parameter values. 

\subsubsection{Search strategy}
We design our search for unstable ECS keeping in mind two groups of objects in phase space: (1) the unstable manifolds of the 1D equilibra and (2) any chaotic attractors. By computing ECS on or near these manifolds, we hope to gain information on nearby time-dependent trajectories. In particular, group 1 is relevant for trajectories that begin near one of the 1D equilibria; such trajectories, at least initially, will approximately follow the equilibrium's unstable manifold.

The Newton-Raphson solver used for solving the FPEs requires good initial guesses to converge to an unstable ECS. Given our goal of using the ECSs to characterize time-dependent trajectories, it is natural to take initial guesses from the latter. For ECSs on or near the unstable manifold of an equilibrium, we take guesses from trajectories that begin near the equilibrium and are perturbed along one of its unstable directions. These are actually the same trajectories defined in section \ref{sec:attractors}, which we used to find attractors. The only difference is that now, we are taking guesses from near the beginning of the trajectory (roughly the first $0 < t < 1000 \tau$) rather than the end. For ECSs on or near chaotic attractors, we use the same set of trajectories, but exclude those that end on an attracting ECS. Our initial guesses are then taken from the asymptotic part of the remaining trajectories.

Depending on the trajectory used to provide initial guesses, the fraction of solver instances that converge can vary from $0$ to $1$. Among those that converge, not all give unique ECSs, so it is essential to detect and remove duplicates. For our case, this is complicated by the fact that, due to invariance with respect to space and time translations, POs and RPOs are 2-dimensional manifolds in the full phase space. In particular, two separate solver instances generally converge to different points on the surfaces, and we need a criterion to detect when two points are on the same surface. To account for this situation, we use the distance measure described in section \ref{subsec:um-shadowing}, which is $0$ if and only if the two ECSs are equivalent up to space and time translations. Thus, if the distance between two ECSs is $0$, they reside on the same `ECS surface', and we remove one of them from our dataset as a duplicate. For a few values of activity, we also ran the Newton-Raphson solver in one or more symmetry subspaces, which helped to identity some of the more unstable ECS. Finally, for many of the ECSs, we performed branch continuation in $\mathrm{R_a}$, which sometimes uncovered ECSs that had been missed by other searches.

 However, none of these searches should be considered exhaustive. In the presence of chaos, there will generally be an infinite number of ECS with arbitrarily large periods. Due to the unpredictable convergence behavior of the Newton-Raphson solver, there is also no guarantee that a given number of searches will uncover all ECS with a given property, such as periods less than some value. A more realistic goal is to design a search strategy that is at least biased toward the most dynamically relevant ECSs, and our method of searching along time-dependent trajectories indeed serves this purpose. Nevertheless, there are other biases that may skew our ECS dataset in different directions. In particular, the Newton-Raphson solver cannot look for ECSs with arbitrarily large periods because it must integrate over at least one period. Such ECSs could be important in some settings, such as near a homoclinic bifurcation. More generally, the dynamical relevance of an ECS is more closely related to its Floquet exponents than its Floquet multipliers \cite{krygier2021exact}. The former describe the local divergence of trajectories near a point on an ECS, whereas the latter quantifies growth or decay of perturbations over an entire period. By contrast, the Newton-Raphson solver is more likely to converge to ECSs with small Floquet multipliers. In future work, we will explore the use of optimization-based methods \cite{farazmand2016adjoint,azimi2022constructing} for finding ECS with larger periods.

\subsubsection{Description and classification}\label{subsec:desc}
ECS data for $\mathrm{R_a} = \{0.1 \, n, n = 1 \ldots 50\}$ can be found in the Supplementary material \cite{suppdata2023}. 
For $\mathrm{R_a} < 0.9$, we documented six equilibria, including the three 1D equilibria in Fig. \ref{fig:1deq}. This number is much smaller than the number of ECS in our datasets at higher activity. While one does indeed expect fewer ECS at small activity, this result may also be influenced by the smaller number of searches performed in this regime. 

Our focus instead was on $\mathrm{R_a} > 0.9$,
where RPOs and higher-dimensional invariant objects begin to dominate the phase space geometry. For each value of $\mathrm{R_a}$ above this threshold, we identified at least 10 ECS. Most of these are RPOs, though POs, TWs, and EQ are also present, as discussed below. For the analysis in section \ref{sec:transient-flows} below, more thorough searches were conducted at a few values of $\mathrm{R_a}$. For example, the $\mathrm{R_a} = 3.4$ dataset (used in Figs. \ref{fig:R34_phasespace_density_shadowing} and \ref{fig:R34_shadowing-examples}) contains 56 ECS, and $\mathrm{R_a} = 4.0$ and $\mathrm{R_a} = 4.5$ have 66 and 210, respectively. 
\begin{figure}
    \includegraphics[width=0.99\textwidth]{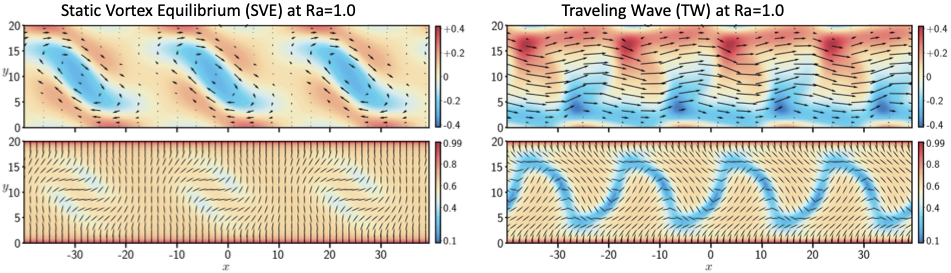}
    \caption{\footnotesize Left: An unstable defectless vortex lattice steady state (equilibrium) with 3-fold translational symmetry $T_3$. Right: An unstable travelling wave with 4-fold  translational symmetry $T_4$. }
        \label{fig:EQ_TW_snapshots}
    \end{figure}
The unstable ECS embody a much larger variety of spatiotemporal patterns compared with the attractors. The discrete translational symmetry $T_n$ is present for $n$ up to 12, and includes two new types of ECS not found among the attractors. The first are defectless, vortex lattice equilibria found by the Newton solver only for relatively small activity ($< 1.5$). As a shorthand, we call these states `static vortex equilibrium' (SVE). These equilibria are not to be confused with stationary vortex (SV) periodic orbits with motile defects discussed earlier. The second are traveling waves; these occur over a wider range of activity, but were found to be unstable in every case. The snapshots of both these types of ECS are shown in Fig. \ref{fig:EQ_TW_snapshots}. The figure `mosaic\_R45.png' in the Supplementary material \cite{suppdata2023} visually expresses the phase space diversity with a mosaic of all 210 ECS found at $\mathrm{R_a} = 4.5$, all of which are unstable.

 \subsubsection{Distribution in phase space}
 	\begin{figure}
    \includegraphics[width=0.49\textwidth]{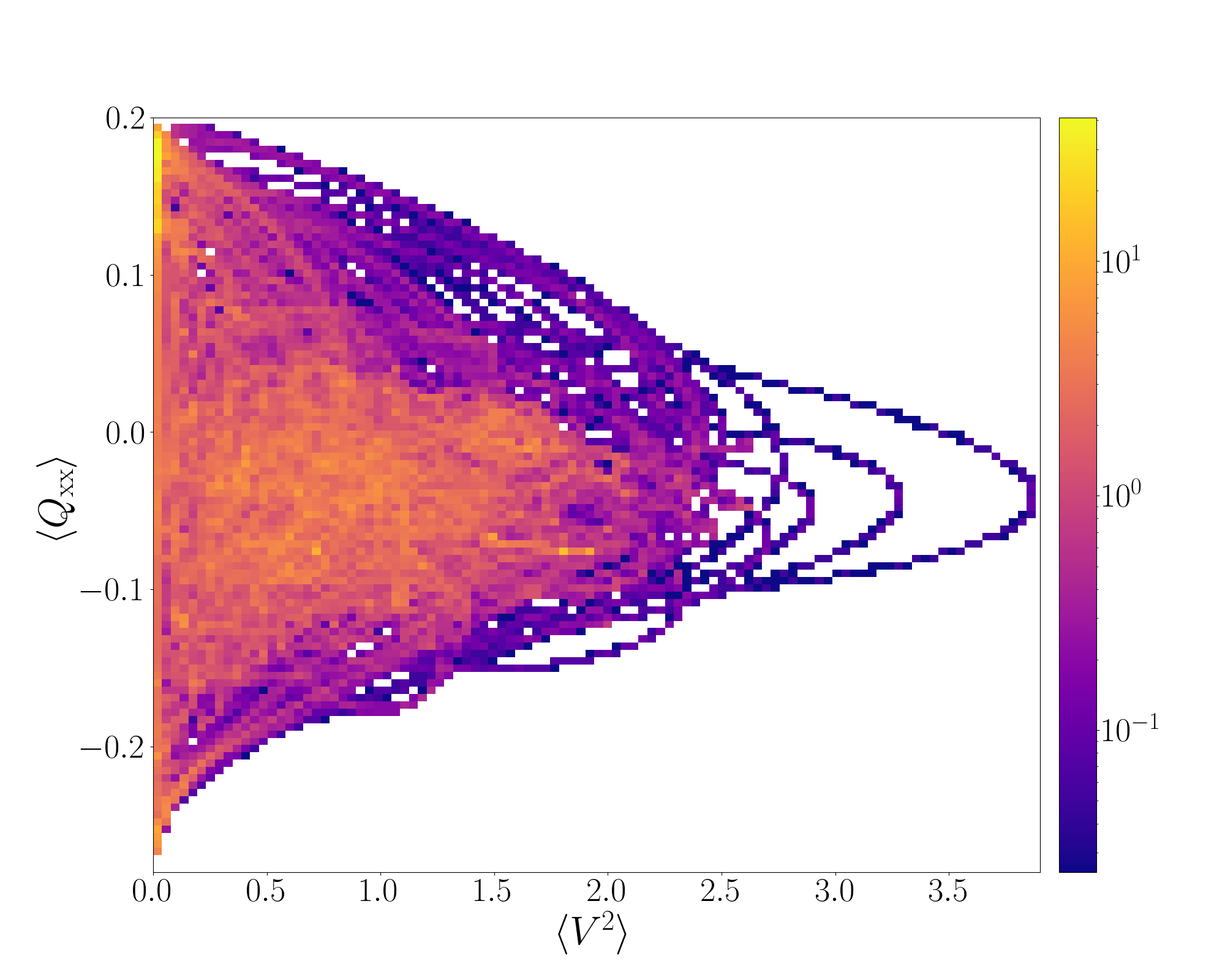}
    \includegraphics[width=0.49\textwidth]{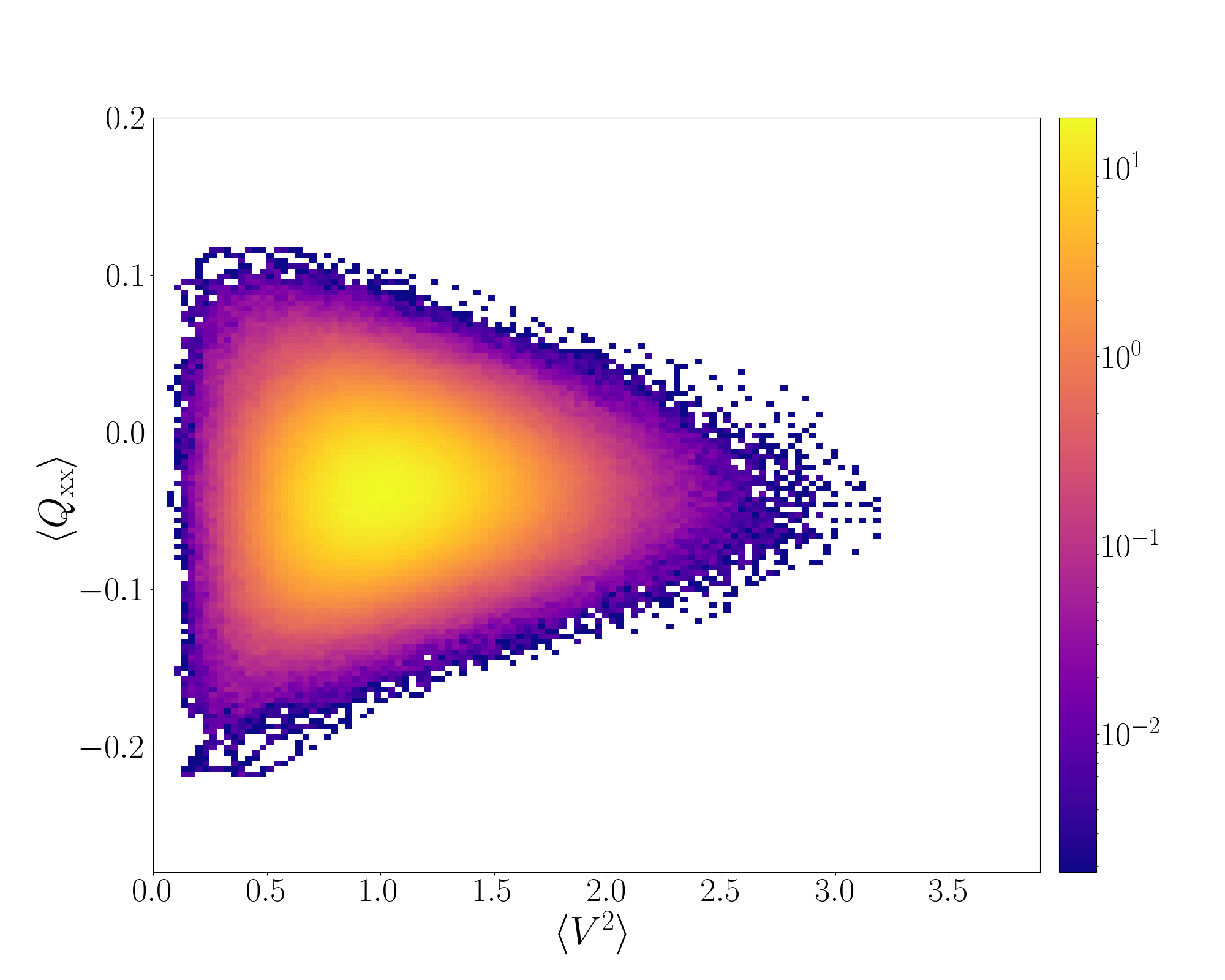}
    \caption{\footnotesize Trajectory `densities' at $\mathrm{R_a}  = 4.5$ for the 201 unstable ECS (left) and turbulent attractor (right), projected on the $(\langle V^2 \rangle,\langle Q_{xx}\rangle)$ plane. Each figure is generated from evenly spaced time series of the corresponding trajectories, which are binned and normalized as a 2D probability density function in $\langle V^2 \rangle$ and $\langle Q_{xx}\rangle$. Trajectories tend to spend more time in the yellow regions versus the violet regions.}
        \label{fig:R45_phasespace_density}
    \end{figure}
 Finally, it is instructive to see the phase space distribution of various ECSs in the turbulent regime. In Fig. \ref{fig:R45_phasespace_density}, we compare the probability density of a long lived turbulent trajectory with that of trajectories on all the computed ECS. The trajectory for each ECS has duration equal to its respective period. In this 2D phase space projection, these ECSs tend to be distributed near the unstable manifolds of the 1D equilibria. In section \ref{sec:transient-flows}, we provide evidence that this conclusion holds in the full phase space as well, i.e., it is not just an accidental feature of the low-dimensional projection in Fig.  \ref{fig:R45_phasespace_density}.
\subsection{Transient flows}\label{sec:transient-flows}
Our interest in unstable flow patterns is driven by their potential to give a reduced-order description of trajectories that are transient or chaotic. By \emph{reduced-order}, we mean theoretical descriptions containing substantially fewer degrees of freedom than the full, time-dependent hydrodynamic equations. Such a description is, by necessity, usually incomplete or approximate, but will at least capture some key features of the dynamics. By \emph{transient trajectories}, we mean finite-time trajectories connecting the neighborhood of an unstable ECS to one of the attractors. For example, one might be interested in the behavior of a system initialized near an unstable equilibrium and subsequently allowed to evolve toward one of the attractors in Fig. \ref{fig:att}. As we demonstrate below, there is substantial variety in the pathways a system might take, so there is no single representative trajectory. Similarly, there is no finite representative trajectory for a turbulent flow; in both cases, the complicated nonlinear dynamics makes it difficult to develop a reduced-order description.

As mentioned in the introduction, the dynamical systems framework \cite{graham2020exact} postulates that such a reduced-order description can be built out of unstable ECS. A recurring motif is a typical turbulent trajectory approaching an unstable, saddle-like ECS along the latter's stable manifold, before departing along the corresponding unstable manifold towards another ECS. There is mounting experimental \cite{crowley2022turbulence} and numerical \cite{krygier2021exact,yalniz2021coarse,page2021revealing,page2022recurrent} in support of this idea for the case of weak inertial turbulence in classical Newtonian fluids. Refs \cite{krygier2021exact,crowley2022turbulence} refined the definition of `close approach' to an ECS, taking into account continuous symmetries and the local geometry in the neighborhood of the ECS. In this way, they defined `shadowing events' as finite time intervals during which a turbulent trajectory and an ECS have quantitatively similar time evolution. By applying this formulation to inertial turbulence, they were able to experimentally detect shadowing events in turbulent Taylor-Couette flow \cite{crowley2022turbulence}. These shadowing events could be used, for instance, as nodes in a network-based model of trajectory statistics. In a complementary work, Ref. \cite{yalniz2021coarse} employed tools from computational topology to construct a Markov chain where the nodes represent neighborhoods of various ECSs, and the transition rates between a pair of nodes are governed by the heteroclinic dynamics between those ECSs. The invariant distribution of the Markov Chain is a coarse grained representation of the probability density function of a generic turbulent trajectory.

Although ECS have been found numerically in non-Newtonian fluid turbulence \cite{page2020exact,dubief2020first}, the usefulness of unstable ECS in such physical settings is still a hypothesis, potentially requiring simplifications such as confinement to small domains and precise experimental control. For active matter systems, these issues have not yet been explored in detail. Here, we provide two pieces of evidence that certain unstable ECS from section \ref{sec:ECS} form a scaffolding of the unstable manifold of UNI, guiding trajectories \emph{en route} from UNI to one of the attractors. Specifically, a typical trajectory from UNI to one of the attractors experiences one or more shadowing events, which could be used as basis elements of a reduced-order model, or as access points for control inputs. Finally, we also report on our attempts to find shadowing in the turbulent regime.

\subsubsection{Trajectory densities in the preturbulent regime} \label{subsec:um-traj-densities}
 \begin{figure}
 \includegraphics[width=0.99\textwidth]{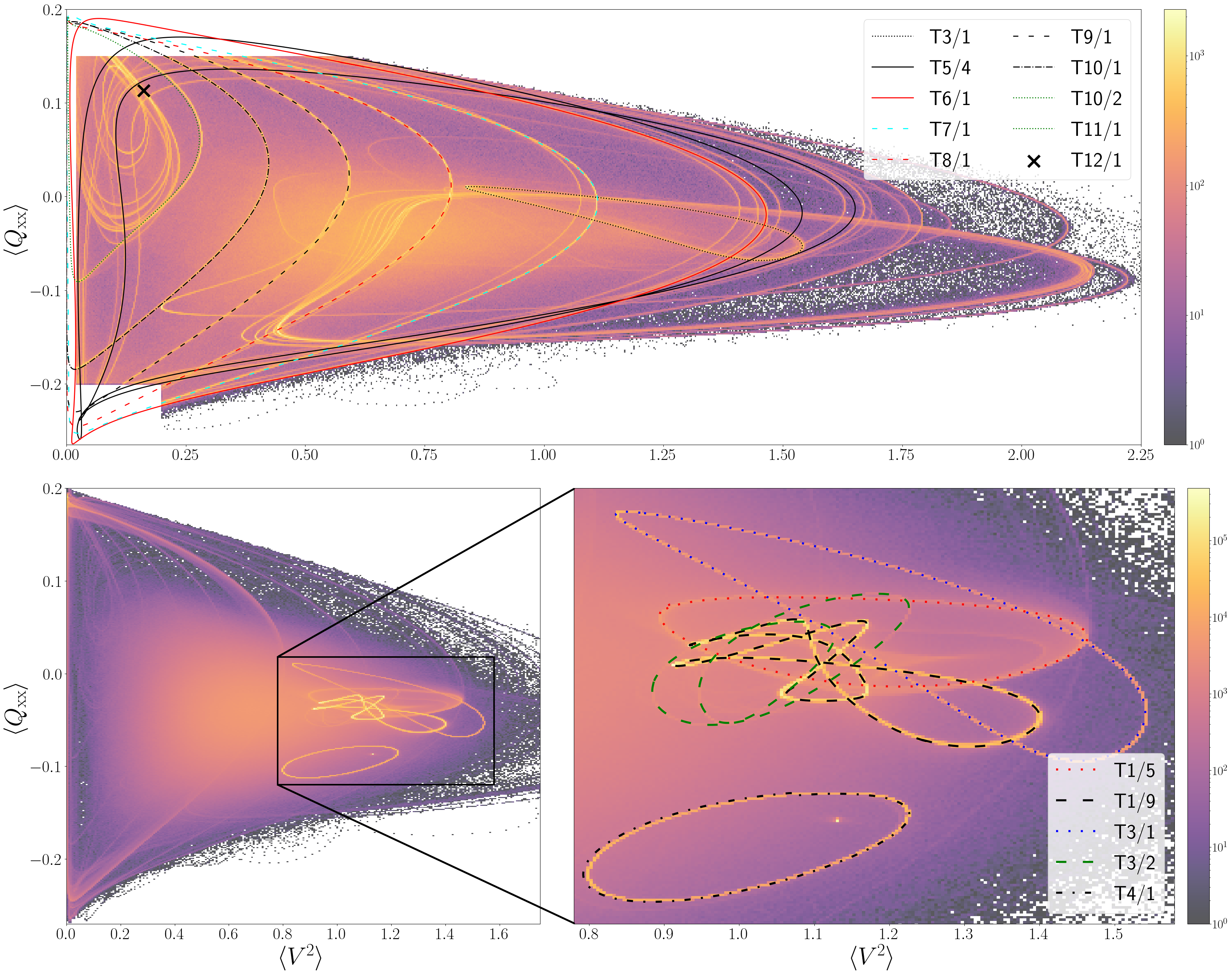}
\caption{\footnotesize Trajectory `densities' along the unstable manifold of UNI in the preturbulent regime projected on the $(\langle V^2 \rangle,\langle Q_{xx}\rangle)$ plane. The dimensionless activity is $\mathrm{R_a} = 3.4$. The full unstable manifold is 34-dimensional and impractical to compute directly. Following the method described in section \ref{subsec:um-traj-densities}, we instead define an ensemble of trajectories that maps out a low-dimensional subspace. The density map (color gradient) is generated from this ensemble by binning evenly spaced time series over the corresponding interval: $0<t<100$ (top) or $100<t<10^4$ (bottom). The apparent overlap of high density regions (yellow) with unstable ECS suggests the occurrence of shadowing, which we have confirmed for selected cases (see section \ref{subsec:um-shadowing}). Note that there are three stable ECS (or attractors): T1/9, T3/1, and T3/2, while the rest of the ECS shown are unstable.}
    \label{fig:R34_phasespace_density_shadowing}
\end{figure}

A simple way to study the unstable manifold of an equilibrium like UNI is to sample it with representative trajectories. These can be generated by perturbing the equilibrium with a linear combination of unstable eigenvectors, which span the tangent space of the unstable manifold in the neighborhood of the equilibrium. Here, we focus on activity $\text{R}_{\text{a}} = 3.4$, where UNI has 34 unstable eigenvectors, and therefore $34-1=33$ coefficients in a general (infinitesimal) perturbation. However, sampling these coefficients uniformly will produce trajectories dominated by the most unstable mode. Hence, to also sample the subdominant directions along the unstable manifold, we construct perturbations from pairs of eigenvectors $\{\mathbf{e}_n,\mathbf{e}_m\}_{n \neq m}$ only. The initial conditions $\mathbf{X}_{\text{init}}$ then take the form:
\begin{equation}
\mathbf{X}_{\text{init}} = \mathbf{X}_{\text{UNI}} + \epsilon \left(\mathbf{e}_n + \alpha \, \mathbf{e}_m \right) \label{eq:UNI-um-ic}
\end{equation}
where $\epsilon = 10^{-5}$, $n \neq m$, and $\alpha$ is a random real number between $-1$ and $1$.
 
In our case, we ran several hundred simulations with these initial conditions and let them evolve from $t=0$ to $t=10^4\tau$. To synthesize the large amount of data that was generated, we constructed time-averaged \emph{trajectory densities} in the 2D phase space projection $(\langle Q_{\text{xx}} \rangle, \langle v^2 \rangle)$. Specifically, we computed a 2D density map by binning the $(\langle Q_{\text{xx}} \rangle, \langle v^2 \rangle)$ time series over a given time interval. Fig. \ref{fig:R34_phasespace_density_shadowing} shows the results for the time intervals $0<t<100 \tau$ and $100 \tau <t<10^4 \tau$. For comparison, the ECS we tabulated at $\mathrm{R_a} = 3.4$ have periods between about $2.6 \tau$ and $22 \tau$. Note that the top panel of \ref{fig:R34_phasespace_density_shadowing} has two rectangular regions cut out from the upper and lower left corners. These regions contain the UNI and ZF equilibria, respectively, and for early times $t < 100 \tau$ are characterized by especially high trajectory density. Hence, to focus on behavior in other regions of phase space, we exclude these areas from the overall density map.

The  trajectory density plot for $0<t<100 \tau$ captures the transient statistics, and regions of high density in that plot are where trajectories spend the most time within the initial transient period. Many of such high density regions are visible in the top panel of Fig \ref{fig:R34_phasespace_density_shadowing}, several of which coincide with unstable (saddle-type) and stable ECS. Nevertheless, other high-density contours (particularly in the top panel corresponding to transients) could not be correlated with ECSs, despite running the Newton-Raphson solver at points along the underlying trajectories. We conjecture that these are unstable 2-tori (quasiperiodic orbits), which are a type of ECS that our current solver is not equipped to compute. The density plot for $100 \tau <t<10^4 \tau$ shown in the bottom panel of Fig. \ref{fig:R34_phasespace_density_shadowing} captures the long term statistics. As expected, the trajectory density is higher on attracting ECS (zoomed-in view in the lower right panel), though strong shadowing of three unstable ECS is also apparent. 

\subsubsection{Shadowing along unstable manifolds in the preturbulent regime} \label{subsec:um-shadowing}
\begin{figure}
    \includegraphics[width=0.99\textwidth]{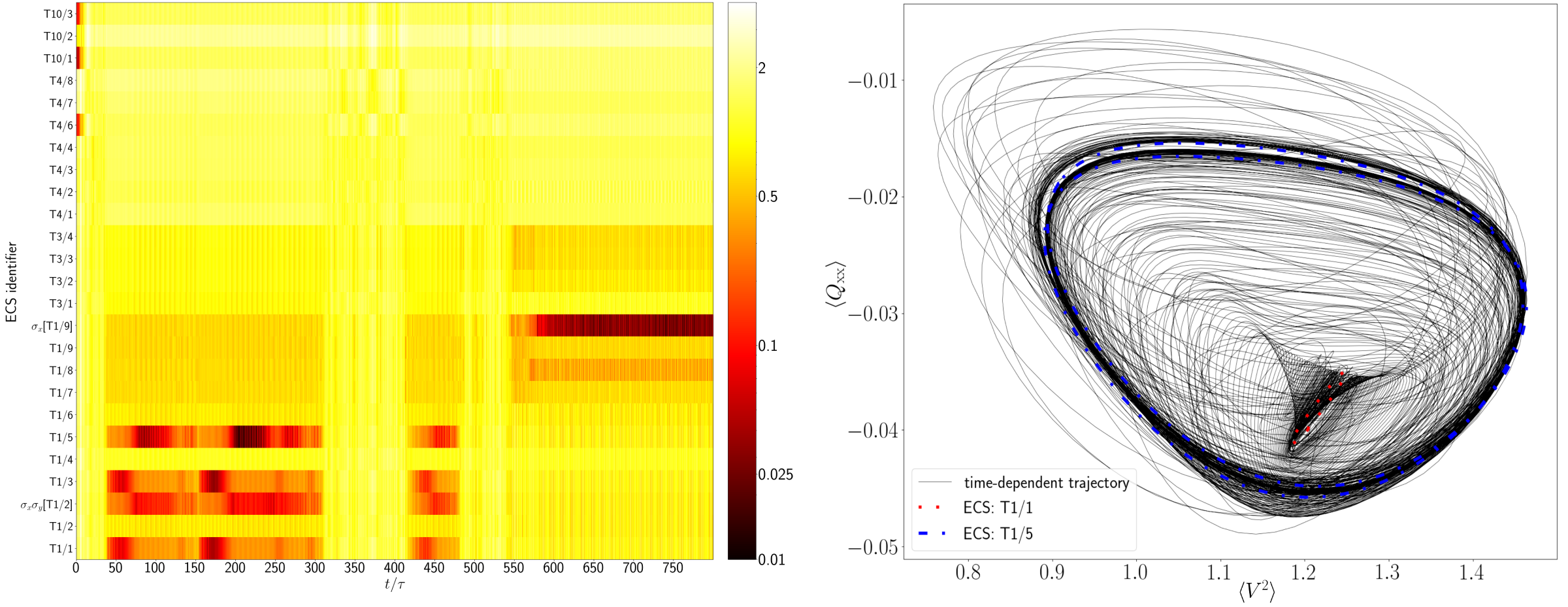}
    \includegraphics[width=0.99\textwidth]{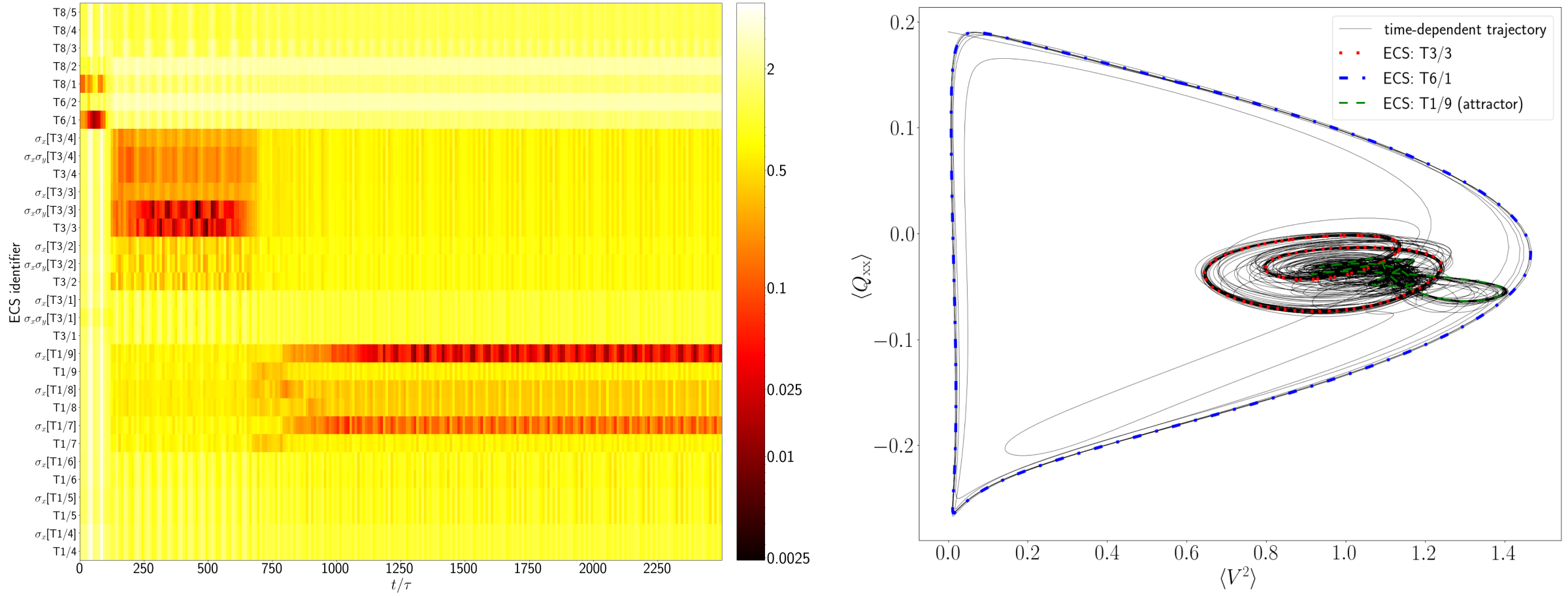}
    \caption{\footnotesize Examples of shadowing along the  unstable manifold of the unidirectional equilibrium in the preturbulent regime at $\mathrm{R_a}=3.4$. The color gradient plots on  the left show the distances (as defined by equation \eqref{eq:symmetry-reduced-distance}) between a time-dependent trajectory and a selection of ECS at the same value of activity. The top left and top right illustrate shadowing of the unstable ECS T1/5 from approximately $200 \tau$ to $250 \tau$. The bottom left and bottom right show two shadowing events: first, the unstable ECS T6/1 from about $30 \tau$ to $70 \tau$, followed by the unstable ECS $\sigma_x \sigma_y (\text{T3/3})$ from $250 \tau$ to $600 \tau$. Both simulations eventually approach T1/9, which is an attractor.}
        \label{fig:R34_shadowing-examples}
    \end{figure}

The second piece of evidence, Fig. \ref{fig:R34_shadowing-examples}, focuses on two of the trajectories used to generate Fig. \ref{fig:R34_phasespace_density_shadowing}. Our aim is to confirm that the \emph{apparent} shadowing of unstable ECS (as suggested by the trajectory densities in the reduced phase space) indeed describes a feature of the full phase space dynamics, and not just an artifact of the low-dimensional projection in Fig. \ref{fig:R34_phasespace_density_shadowing}. To do so, it is necessary to define a measure of distance between two trajectories in phase space. The natural metric is simply the $L^2$ norm of the fields. However, due to the continuous translational symmetry in the streamwise direction ($x$ coordinate), this metric is inadequate for identifying \emph{dynamical} similarity between trajectories. For example, the $L^2$ norm may assign a large distance between two trajectories that evolve nearly identically, but only up to a constant streamwise translation. To account for this situation, we use the following time-dependent continuous-symmetry-reduced distance between a trajectory $X_1(t)$ and an ECS $X_E(t)
$:
\begin{equation}
d_{1E}(t)=\min_{\substack{0 \leq l \leq L \\ 0\leq\tau\leq T}}\|\tau_x(l)(X_1(t))-X_E(\tau)\|_2,\label{eq:symmetry-reduced-distance}
\end{equation}
where $T$ is the time-period of the ECS ($T=0$ if it is an equilibria or a travelling wave).
The top left and bottom left panels in Fig. \ref{fig:R34_shadowing-examples} show the results of applying this measure to two trajectories initialized from Eq. \ref{eq:UNI-um-ic}. The symmetry-reduced distance (color gradient) is computed at a set of fixed timepoints (horizontal axis) and with respect to several ECSs (vertical axis). Several shadowing events are clearly visible as horizontal red/black segments, where the symmetry-reduced distance is small. The corresponding reduced phase space trajectories are shown on the right. The full time series data of distances of both trajectories from all the ECS is provided in the Supplementary material \cite{suppdata2023}.
  
\subsubsection{Testing for shadowing in the turbulent regime} \label{subsec:um-shadowingturb}
To test whether the turbulent state shadows any ECS, we consider typical trajectories in the turbulent regime at $\mathrm{R_a} = 4.5$. As discussed in Section \ref{subsec:desc}, we computed $210$ distinct unstable ECS at this parameter value. Each ECS $X_E(t)$ has three discrete-symmetry related counterparts, i.e., $\sigma_x(X_E(t)), \sigma_y(X_E(t))$ and $\sigma_y\sigma_x(X_E(t))$. We computed continuous-symmetry-reduced distances using Eq. \ref{eq:symmetry-reduced-distance} between each of the the 840 ECS, and the 22 trajectories starting on the unstable manifold of UNI. These trajectories are obtained by picking initial conditions that are small perturbations of UNI along each of its 22 unstable directions. 

Unlike the preturbulent regime, we found no instances of shadowing of any of the ECS by the turbulent trajectories. Although there are some instances of close approaches to certain ECS, the evolution of the chaotic trajectory does not resemble any of the ECS even for times much smaller than the typical time period of the corresponding ECS \cite{krygier2021exact}. 
\begin{figure}[h!]
    \includegraphics[width=0.45\textwidth]{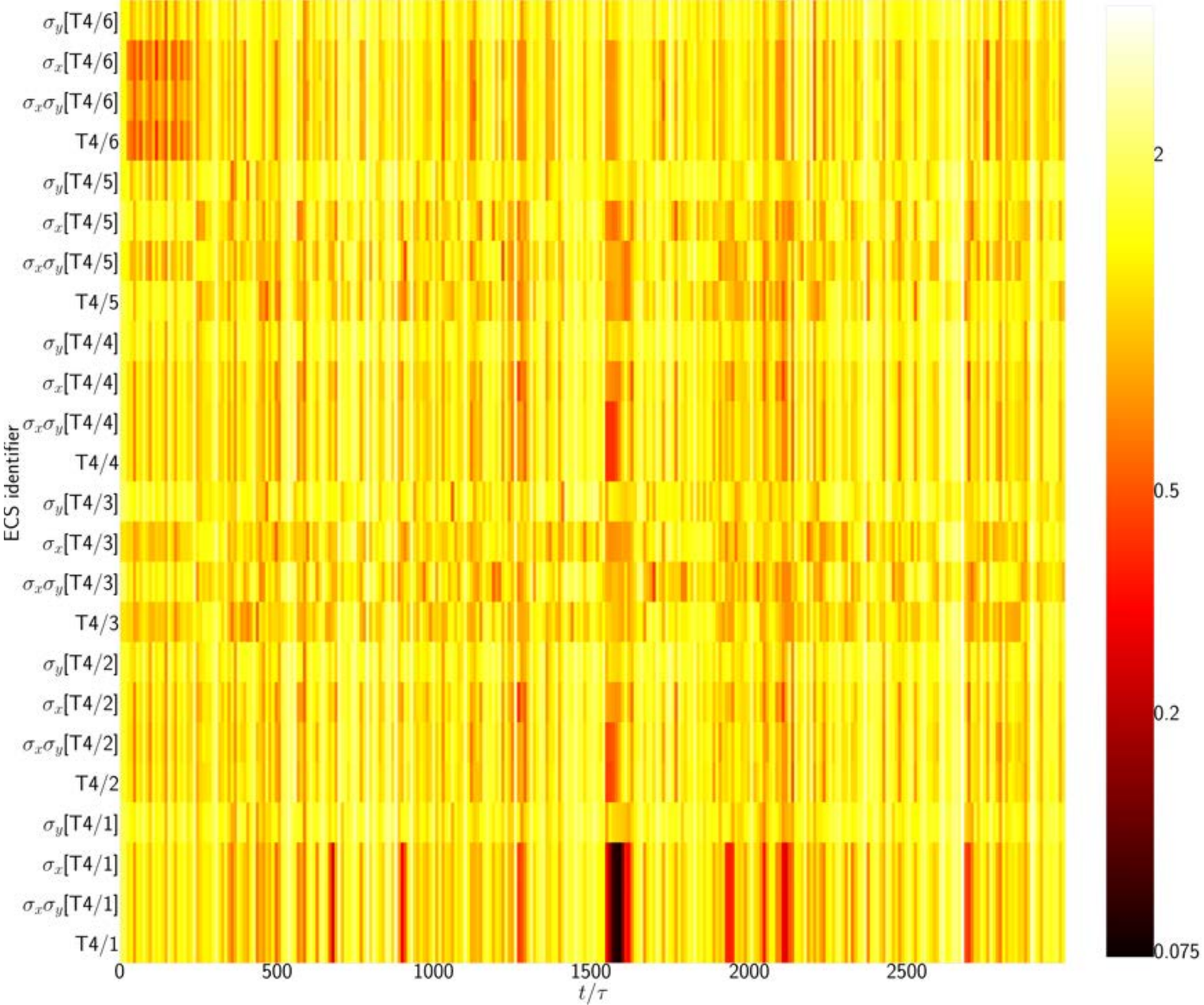}
    \includegraphics[width=0.54\textwidth]{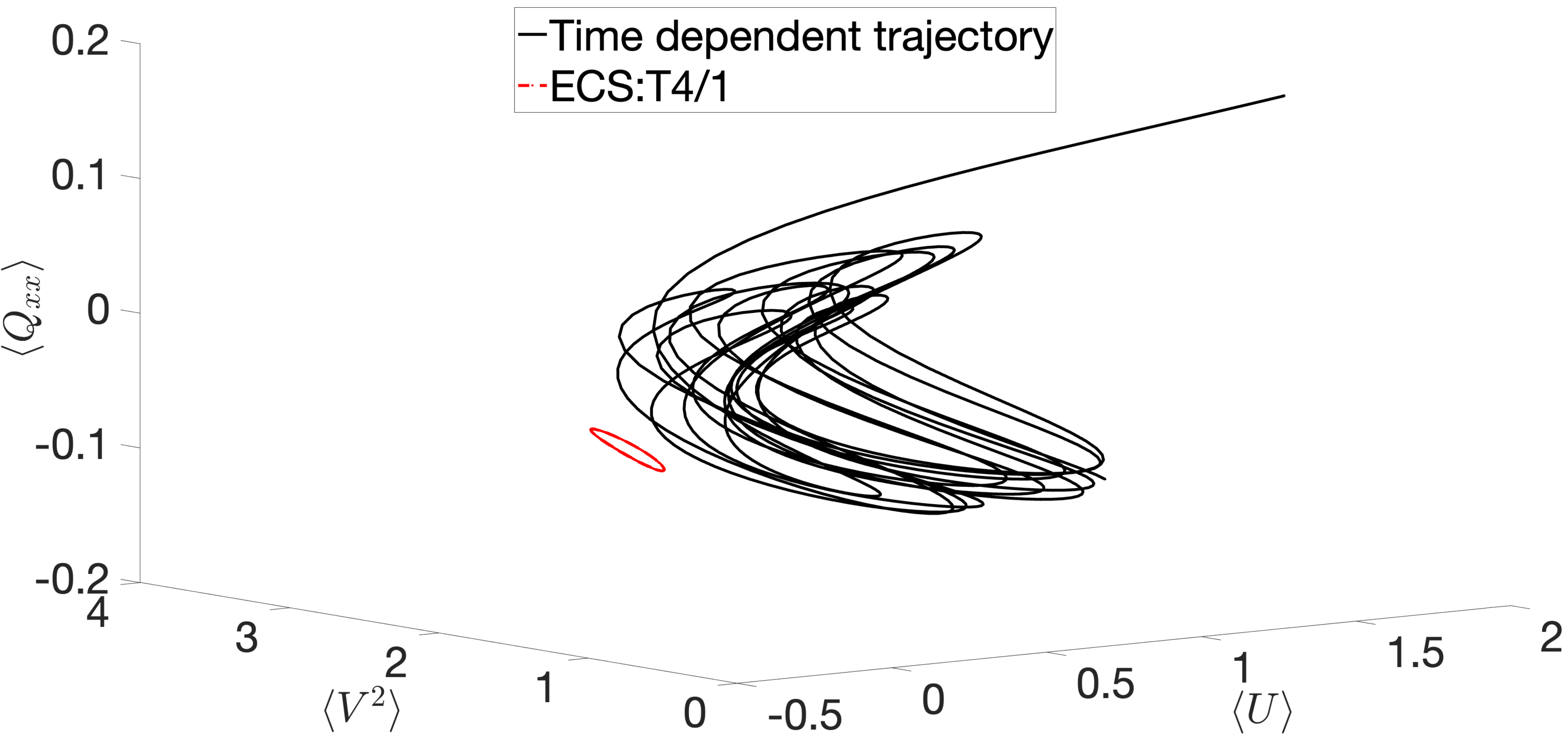}
    \caption{\footnotesize Left: A subset of the distance plot chosen to highlight one of the closest encounters between a chaotic trajectory and an ECS in the turbulent regime at $\mathrm{R_a}=4.5$. The trajectory visits the neighborhood of an ECS (labeled T4/1) at $t/\tau\approx 1500$. Right: The trajectory from  $t/\tau= \text{0 to } t/\tau\approx 1800$ in black along with the ECS T4/1. There is clearly no shadowing of the ECS, and the trajectory appears to be quasiperiodic. The full time series data of distances of this trajectory from all the ECS is provided in the Supplementary material \cite{suppdata2023}.}
    \label{fig:quasi1}
    \end{figure}
 By visual inspection of the 3D phase space plot, we found that some segments of the chaotic trajectories appear to be quasiperiodic, see the right panel of Fig. \ref{fig:quasi1} for one example. This leads us to conjecture that these trajectories might be shadowing quasiperiodic ECS of the system. Recall that the trajectory density plot shown Fig. \ref{fig:R45_phasespace_density} of the preturbulent regime was also suggestive of shadowing of quasiperiodic ECS. 
 
 Most prior work on ECS in turbulent fluids has been limited to computing equilibria, POs and RPOs \cite{suri2019heteroclinic}. Two recent works \cite{parker2022invariant, parker2023predicting} have highlighted the importance of quasiperiodic (QPO) ECS in determining the statistics of chaotic dynamical systems. A quasiperiodic trajectory with n incommensurate frequencies densely covers a n-tori in the phase space. The authors argue that since n-tori are structurally unstable for $n>2$, equilibria, POs, RPOs and 2-tori form the complete set of invariant objects for understanding the phase space transport in chaotic systems. These two works have proposed efficient algorithms for computing 2-tori, and also demostrated that approximate statistics of the chaotic forced Kuramoto-Sivashinsky PDE can be captured using a modified periodic orbit theory adapted to QPOs. 

Our set of ECS could also be missing some POs/RPOs that lie in the region where the local curvature of the chaotic attractor is too high for the Newton-Rhapson hookstep algorithm to converge. The computation of QPO ECS in the AN channel flow, as well as further work on shadowing of ECS in turbulent active nematics will be taken up in a future publication.

\section{Discussion and Concluding Remarks}
We have used the ECS framework to provide a comprehensive study of the stable and unstable steady states of the active nematic system in a 2D periodic channel. We found that in the preturbulent regime, multiple regular attractors co-exist with a chaotic attractor. This implies that a subcritical transition to chaotic flow is possible in such systems. This also hints at the presence of an edge manifold in the phase space, separating the regular states from the chaotic ones. 

Our attractor computations confirm that as activity is increased, the general sequence of stable states i.e., zero flow, 1D flow, vortex lattice, followed by transition to turbulence, identified in previous studies \cite{shendruk2017dancing, doostmohammadi2019coherent,thampi2022channel}, is correct. However, visual inspection is not enough to identify dynamic similarity of various flow states, and phase space analysis aided by symmetry considerations provide additional insight. In addition to the above mentioned attractors, our computations revealed the existence of counter-rotating lattice vortex lattice, traveling wave and modulated wave attractors. 

We demonstrated that unstable ECS organize the short term dynamics of perturbations around the unstable unidirectional steady state in the preturbulent regime. Typical trajectories lying on the unstable manifold of the unidirectional equilibrium were found to shadow a subset of the unstable ECS in the transient phase, and eventually converge onto one of the attractors in steady state.

In the turbulent regime, which is characterized by the lack of any attractors, we found over 200 unstable ECS at a single set of parameters. Low dimensional projections suggest that these ECS exist in the same region of phase space as the typical turbulent trajectories of the system. However, unlike the preturbulent regime, there was no evidence of shadowing of the PO/RPO/TW type ECS by the turbulent trajectories, and we conjecture that these trajectories are instead shadowing quasiperiodic ECS. 

Future work will focus on computing quasiperiodic ECS as well as PO/RPOs of larger periods using extensions of our existing Newton-Rhapson solver, or optimization-based methods. It would also be interesting to compare the ECS discovered in the current work to dynamic modes obtained in active turbulence using data-driven dynamical systems approaches such as Proper Orthogonal Decomposition and Dynamic Mode Decomposition \cite{henshaw2023dynamic}. Finally, detection of stable and unstable ECS in an experimentally realizable active nematic system such as the 2D annulus \cite{chen2018dynamics,zarei2020role,joshi2023dynamics} will be considered. 

 \begin{acknowledgments}
This material is based upon work supported by the U.S. Department of Energy, Office of Science, Office of Basic Energy Sciences under award number DE-SC0022280. This work was completed utilizing the Holland Computing Center of the University of Nebraska, which receives support from the Nebraska Research Initiative. We thank Aparna Baskaran, Seth Fraden, Mike Hagan, Chaitanya Joshi, and Zahra Zarei for helpful discussions.
\end{acknowledgments}

\bibliography{refs}
\end{document}